\documentclass[aps,prl,a4paper,preprint,superscriptaddress]{revtex4}
\usepackage[utf8]{inputenc} 
\usepackage{graphicx}
\usepackage{amsmath,epsfig,amssymb}
\usepackage{color}

\def\be{\begin{equation}}
\def\ee{\end{equation}}
\def\bea{\begin{eqnarray}}
\def\eea{\end{eqnarray}}

\begin{document}

\title{All-Optical Generation of Surface Plasmons in Graphene}
\date{\today}

\author{T.~J.~Constant}
\email{t.j.constant@ex.ac.uk}
\affiliation{Electromagnetic Materials Group, Department of Physics, College of Engineering, Mathematics and Physical Sciences, University of Exeter, Exeter, Devon, UK. EX4 4QL.}

\author{S.~M.~Hornett}
\affiliation{Electromagnetic Materials Group, Department of Physics, College of Engineering, Mathematics and Physical Sciences, University of Exeter, Exeter, Devon, UK. EX4 4QL.}

\author{D.~E.~Chang}
\email{darrick.chang@icfo.es}
\affiliation{ICFO - Institut de Ci\`encies Fot\`oniques, Mediterranean
Technology Park, 08860 Castelldefels (Barcelona), Spain}

\author{E.~Hendry}
\affiliation{Electromagnetic Materials Group, Department of Physics, College of Engineering, Mathematics and Physical Sciences, University of Exeter, Exeter, Devon, UK. EX4 4QL.}

\maketitle
\textbf{Surface plasmons in graphene offer a compelling route to many useful photonic technologies \cite{Grigorenko2012,Abajo2014,koppensnl}. As a plasmonic material, graphene offers several intriguing properties, such as excellent electro-optic tunability\cite{Craciun2011}, crystalline stability, large optical nonlinearities\cite{Hendry2010a} and extremely high electromagnetic field concentration\cite{Brar2013}. As such, recent demonstrations of surface plasmon excitation in graphene using near-field scattering of infrared light\cite{Fei2012,Chen2012} have received intense interest. Here we present an all-optical plasmon coupling scheme which takes advantage of the intrinsic nonlinear optical response of graphene. Free-space, visible light pulses are used to generate surface plasmons in a planar graphene sheet using difference frequency wave mixing to match both the wavevector and energy of the surface wave. By carefully controlling the phase-matching conditions, we show that one can excite surface plasmons with a defined wavevector and direction across a large frequency range, with an estimated photon efficiency in our experiments approaching $10^{-5}$.}

Graphene has attracted significant interest in recent years as a unique optical material. In particular, it has been predicted and experimentally shown that graphene can support very highly confined surface plasmon modes\cite{Grigorenko2012,Bludov2013}, with an electrically tunable dispersion\cite{Fei2012,Chen2012}. Despite these promising discoveries, the burgeoning field of graphene plasmonics has some serious obstacles to overcome if it is to progress from the proof-of-principle stage. Problems arise due to the small wavelength of the surface plasmons, two orders of magnitude smaller than light of the same frequency. This has led to the development of specialised measurement techniques, most of which use infrared light and geometries with scattering resonances \cite{Gao2012a,Alonso-Gonzalez2014,Gao2013} or near-field sources \cite{Fei2012,Chen2012} to excite graphene surface plasmons. However, the far-infrared regime, in which graphene plasmons are predicted to have long lifetimes, lacks developed sources and detectors compared to the visible regime. Alternative approaches, such as the manipulation of surface acoustic waves to couple to the graphene surface plasmons \cite{Farhat2013,Schiefele2013}, therefore hold promise.

Particularity desirable is the potential to excite a plasmon eigenstate with a singular energy, momentum and direction, vital for many future applications, including plasmonic circuits. In this respect, very recent progress has been made, with the development of carefully designed nano antennas which can locally excite and direct surface plasmons in graphene \citep{Alonso-Gonzalez2014}. Here, the combination of infrared source frequency and nanoantenna dimensions determine the frequency, wavevector and direction of the surface plasmons generated. In this article, we investigate a competing approach that embodies many of these desirable aspects of directivity without requiring careful nanofabrication of antennas. This all-optical approach can access a distinctly broad frequency range, even down to the far infrared. We coherently excite surface plasmons using two visible frequency free-space beams via difference frequency generation (DFG), an effect which we monitor through changes in reflectance, and can tune the frequency and wavevector of the surface plasmon through careful adjustment of incident light sources. This potential to excite and detect plasmons purely with free space optics, and at frequencies different than that of the plasmons themselves, has the potential to significantly expand the technological possibilities for graphene plasmonics.

\begin{figure}[b]
\begin{center}
\includegraphics[width=12cm]{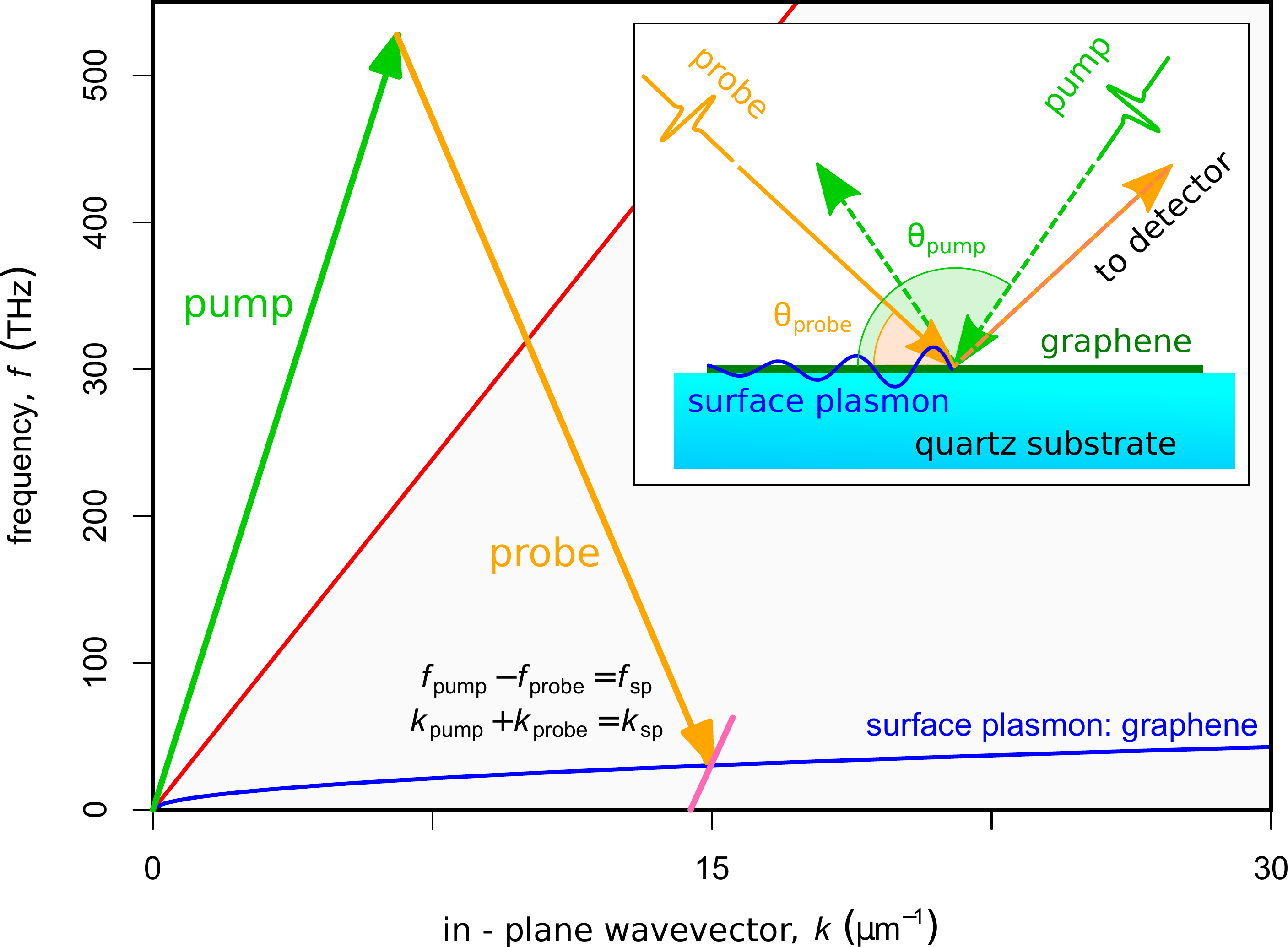}
\end{center}
\caption{The nonlinear coupling scheme illustrated on a dispersion diagram. The DFG of the pump (green arrow) and probe (orange arrow) allows access to wavevectors outside of the light line (red line). This permits phase-matching to the surface plasmon modes in graphene (blue line). The pink line illustrates a region that can be interrogated by altering the pump wavelength from $615~\mathrm{nm}$ to $545~\mathrm{nm}$ with the probe wavelength fixed at $615~\mathrm{nm}$. (Inset) The experimental arrangement used to excite surface plasmons on graphene.\label{fig:coupling}}
\end{figure}

The intrinsic nonlinear interactions of graphene with light are surprisingly large\cite{Hendry2010a,Manzoni,Dean2010,Hong2013,Gu2012}. Moreover, large enhancements of nonlinear optical effects are predicted by the presence of highly confined plasmons in graphene\cite{Gullans2013,Mikhailov2011}. It seems intuitive, then, to attempt the converse: to use the nonlinear interaction between optical fields to resonantly drive surface plasmons. This kind of approach has been demonstrated for thin metallic films \cite{Renger2009,Palomba2008}, and has been recently proposed for graphene, with various coupling schemes for the difference frequency mixing of infra-red light in graphene clad waveguide structures suggested \cite{Yao2014,Sun2014}. Similar in concept, figure 1 shows our nonlinear coupling scheme illustrated on a dispersion diagram. By illuminating the graphene with two intense laser pulses with well-defined angles of incidence but different frequency, labeled here $f_{pump}$ and $f_{probe}$, one can phase match both the frequency and wavevector, $k$, of the surface plasmon. This wave mixing process is a second order nonlinear effect, normally forbidden in centro-symmetric crystals\cite{Boyd2008}, but possible in graphene because of the distinctively non-local, spatial character of the interaction \cite{Mikhailov2011}. The inset in fig. 1 shows the experimental arrangement used. An identical pair of optical parametric amplifiers (OPAs), pumped by an amplified femtosecond laser system, generate the ~100 fs-pulses at a repetition rate of 1 kHz. The wavelengths of the two OPAs are selected independently, and the beams are directed to the sample. The experiments are carried out in a non-collinear geometry, using two beams incident on the samples at angles $\theta_{pump}$ and $\theta_{probe}$. Oblique angles provide sufficient in-plane momentum to match to the surface plasmon, as illustrated in fig. 1. The incident beams are weakly focused on the sample using 30 cm focal length lenses, giving rise to a very small uncertainty in angle $\sim0.017$ rad, and a similarly negligible uncertainty for the in-plane wavevectors. Sets of half-waveplates and polarizers determine both the average power and polarization, with the polarization set such that the electric vector of the light is in the plane of incidence (transverse magnetic polarized).
The pump pulse fluence, $\Phi$, used is typically in the range $\Phi\sim0.1-0.2~\mathrm{mJ/cm^2}$, with a pump spot size on the sample of $\sim 300~\mathrm{\mu m}$ radius. This pump fluence is an order of magnitude less than the photo-modification threshold for graphene\cite{Alexeev2013}, and the probe fluence is typically two orders of magnitude smaller still. We measure the differential reflection of the probe beam defined as $\Delta R / R=(R-R_0)/R_0$ where $R$ and $R_0$ are the reflections with and without the presence
of the pump pulse, respectively. In order to isolate the nonlinear reflection signal, we vary the temporal overlap of the two pulses using a motorized delay stage.

\begin{figure}[b]
\begin{center}
\includegraphics[width=12cm]{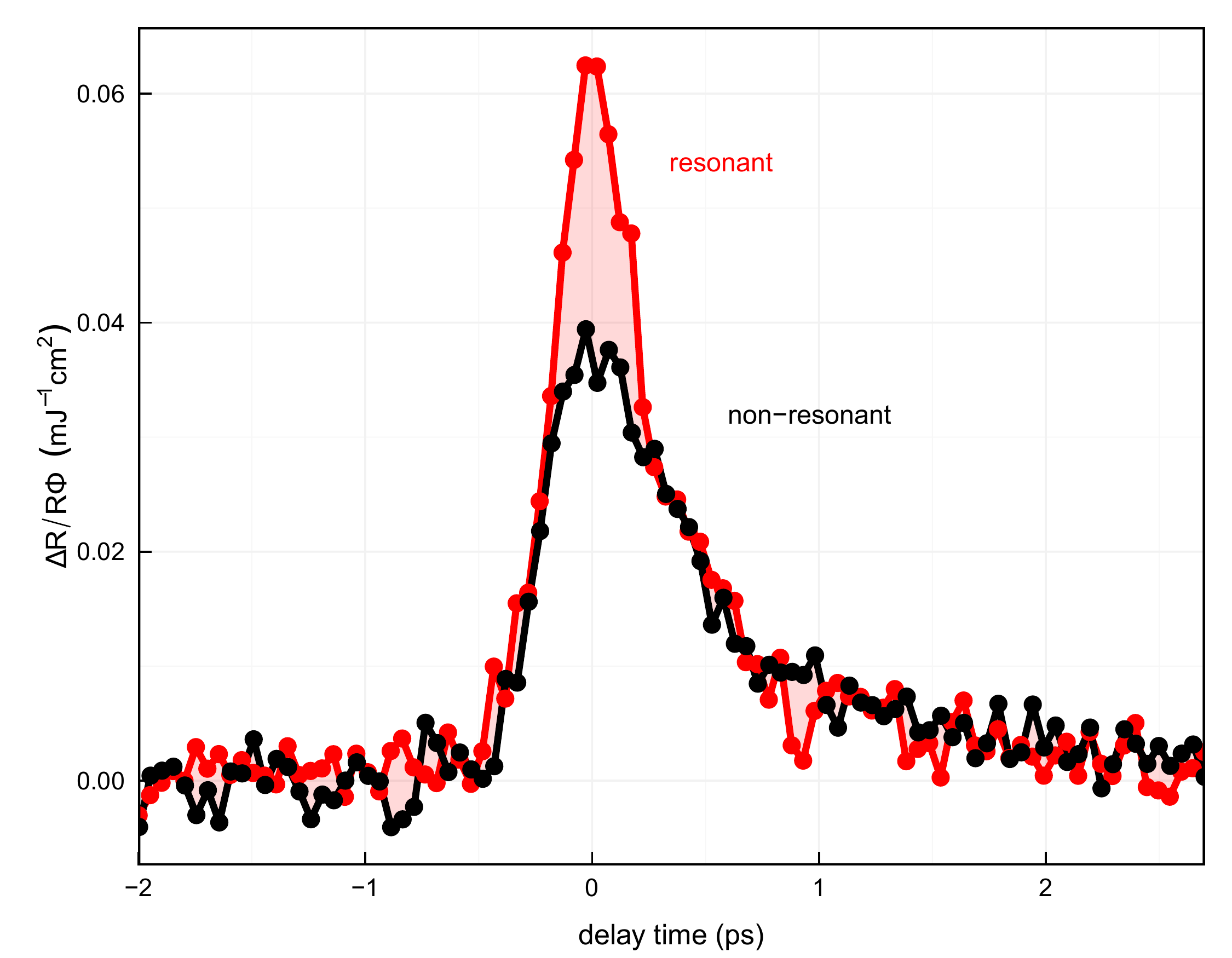}
\end{center}
\caption{Differential reflection normalized to average pump fluence as a function of temporal overlap for the geometry $\theta_{pump} = 15^\circ$, $\theta_{probe} = 125^\circ$. At zero delay time, both the pump and probe pulses arrive simultaneously, leading to a nonlinear change in the probe reflection. Two curves are shown: The black curve labelled ``non-resonant" shows a typical time asymmetric mesurement when the difference frequency produced by the pump and probe ($61.2~\mathrm{THz}$) does not coincide with a surface plasmon energy state. The red curve shows an addittional fast symmetric contribution to the recorded reflection signal when the differency frequency matches the energy of a graphene surface plasmon ($23.8~\mathrm{THz}$).\label{fig:example_pump_probe}}
\end{figure}

For optical excitation pulses, one expects optical nonlinearity arising due to saturable absorption caused by Pauli blocking of interband transitions\cite{Brida2013}. A typical measurement of the temporal dynamics recorded for this process ($\lambda_{pump} = 547~\mathrm{nm},~\lambda_{probe} = 615~\mathrm{nm}$) is shown by the black curve in fig.~\ref{fig:example_pump_probe}. The asymmetric line shape of the signal is due to the relaxation dynamics of the excited electrons cooling\cite{Dawlaty2008,Hale2011b}, with significant temporal broadening caused by the spatial overlap of the non-collinear beam spots. Note that there is no appreciable signal from the quartz substrate (See Supplementary Information (SI) fig.~\ref{fig:substrate_check}).

For non-degenerate pump and probe beams, in addition to (incoherent) saturable absorption effects, one can expect (coherent) wave mixing signals. This coherent contribution to the probe reflection is expected to be significantly enhanced when the difference frequency field generated by the pump and probe matches that of the graphene surface plasmons. This is analogous to that of a stimulated Raman process, corresponding to a transfer of energy from pump to probe pulses\cite{Boyd2008} via the generation of surface plasmons. An example of the recorded temporal dynamics under such a resonant condition is presented in  fig.~\ref{fig:example_pump_probe}.
%
%
Comparing the two curves in this figure, we see that the ``non resonant" signal (i.e. when one is not phase matching to plasmon excitation) gives rise to an asymmetric lineshape representative of carrier cooling dynamics. Under ``resonant" conditions (i.e. when phase matching conditions are satisfied) we observe a fast additional contribution to the signal, giving rise to a more symmetric lineshape, as one would expect for a coherent signal. For certain experimental geometries and excitation fluences, signal enhancements of up to approximately $\times 4$ can be observed (see supplementary figure \ref{fig:low_fluence}). It should be noted that, depending on efficiencies, it may be possible to isolate the coherent signal using a heterodyne detection scheme\cite{Borri2000}, which could also allow detection of a plasmon in a different spatial position than generated. 

In order to observe the presence of the coherent signal, we vary the difference frequency in order to isolate any resonant, coherent conditions. In experiment, the pump wavelength is varied from $615~\mathrm{nm}$ to $545~\mathrm{nm}$, with the probe wavelength set at $615~\mathrm{nm}$, allowing difference frequencies ranging from 0 to 60~THz. In this way, it is possible to interrogate a section of the surface plasmon dispersion, for example the region illustrated by the pink line in fig.~\ref{fig:coupling}.  Note that we normalize the signal by pump fluence in order to remove artefacts due to power variation\cite{Dawlaty2008}. By altering the experimental geometry, we investigate here three different regions of the dispersion diagram corresponding to ($\theta_{pump} = 55^\circ, \theta_{probe} = 45^\circ$), ($\theta_{pump} = 50^\circ, \theta_{probe} = 70^\circ$) and  ($\theta_{pump} = 15^\circ, \theta_{probe} = 125^\circ$).
\begin{figure}[b]
\begin{center}
\includegraphics[width=15cm]{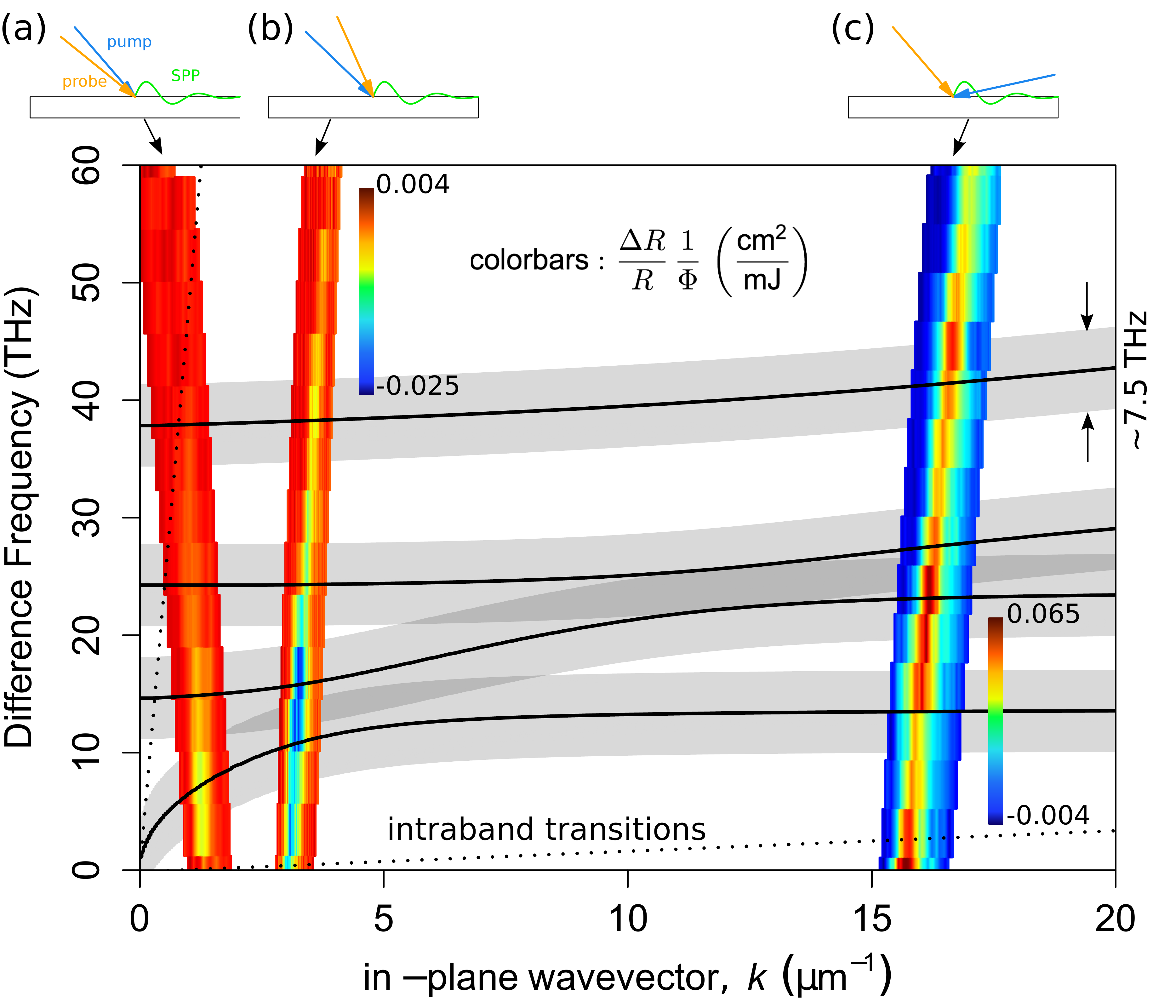}
\end{center}
\caption{Plots of differential reflection normalized to pump fluence for the three different experimental geometries, superimposed on the graphene surface plasmon-phonon dispersion. (black lines). Angles used are: (a) $\theta_{pump} = 55^\circ, \theta_{probe} = 45^\circ$, (b) $\theta_{pump} = 50^\circ, \theta_{probe} = 70^\circ$ and  (c) $\theta_{pump} = 15^\circ$, $\theta_{probe} = 125^\circ$. The intraband transition threshold and light line (dotted lines) are labelled on the diagram. The grey regions indicate the spectral resolution of our experiment.\label{fig:dispersion_compilation}}
\end{figure}

Figure \ref{fig:dispersion_compilation} shows the results of these three measurement geometries, superimposed on the surface plasmon dispersion (black line). The dispersion was calculated according to the model outlined in ref.~\cite{Luxmoore2014}, with the $\mathrm{SiO_2}$ substrate phonon frequencies as given and a Fermi energy of $E_f=0.5~\mathrm{eV}$. This Fermi energy is larger than the expected intrinsic doping of our graphene samples (see Methods for sample details), which we attribute to a significantly raised electron temperature expected under illumination by ultrafast pulses (see SI, fig.~\ref{fig:high_fluence}). Hybridization with the substrate phonons leads to four branches\cite{Luxmoore2014,Yan2013}. The overlaid colour plots are placed on the diagram so that the maximum differential reflection signal achieved in each delay-scan corresponds to the difference frequency and wavevector of the dataset. The grey shading around the plasmon dispersion curve indicates the expected spectral broadening of the signals ($\sim 7.5~\mathrm{THz}$) due to the finite bandwidth of $\sim 100$~fs-pulses.

Near the regions defined by the surface plasmon dispersion in graphene, we observe clear enhancement in the differential reflection.  The assignment of the spectral features to surface plasmon excitation is further supported by the polarization dependence of the signal (see fig.~\ref{fig:polarisation}). The observation of these resonant features over the incoherent background is strongly dependent on the magnitudes of both pump and probe intensities (see SI figures \ref{fig:high_fluence} and \ref{fig:low_fluence}). For larger difference frequencies, up to 150~THz, we do not observe any further resonance features in our spectra (see fig.~\ref{fig:high_freqs}). 

The lower branch of the plasmon dispersion relation gives rise to the largest mixing signals for the low wavevector phase-matching (angles (a) and (b) in fig.~\ref{fig:dispersion_compilation}), while the upper frequency branches give rise to largest signals for the high wavevector (angle (c) in fig.~\ref{fig:dispersion_compilation}) region. Whilst we observe clear resonance features in all three of these experimental geometries, we also observe a change in sign of the signal between low (figure \ref{fig:dispersion_compilation}(a) and (b)) and high (figure \ref{fig:dispersion_compilation}(c)) wavevector regions. 
%
%
The absolute differential reflectivity signal size also increases with increasing wavevector.
%
%

To understand the origin of these coupling behaviours, we have developed a simple theoretical model that captures the salient features of this nonlinear reflection and generation of plasmons, and we briefly summarize the ideas here (details of the model are presented in the supplementary material). In general, equations for the electromagnetic boundary conditions at the air-graphene-substrate interface relate the wavevector and frequency dependent reflection and transmission coefficients $r(k,\omega), t(k,\omega)$ to the graphene current density $J(k,\omega)$. The current density, on the other hand, can be written in terms of the electric field via conductivity functions, which allows the equations to be solved in terms of fields alone. Nonlinear contributions imply that $J(k,\omega)$ depends on fields at other wavevectors and frequencies, which couple the various reflection and transmission coefficients together. For a second-order conductivity $\sigma^{(2)}$, we find that the probe transmission depends on the pump via
\be
t_{probe}=\frac{t_{probe}^{(L)}}{1-A\left(\sigma^{(2)}\right)^2 |t_{pump} |^2 \, I_{pump}},\label{eq:tprobemain}
\ee
with an analogous equation for the pump transmission (expressions for $r$ are more involved but are directly related to $t$, see supplementary materials). Here $t_{probe}^{(L)}$ is the linear transmission coefficient, A is a function of linear optical properties and beam angles, and for notational simplicity, dependencies on $k,\omega$ are implicit.
\begin{figure}[b]
\begin{center}
\includegraphics[width=15cm]{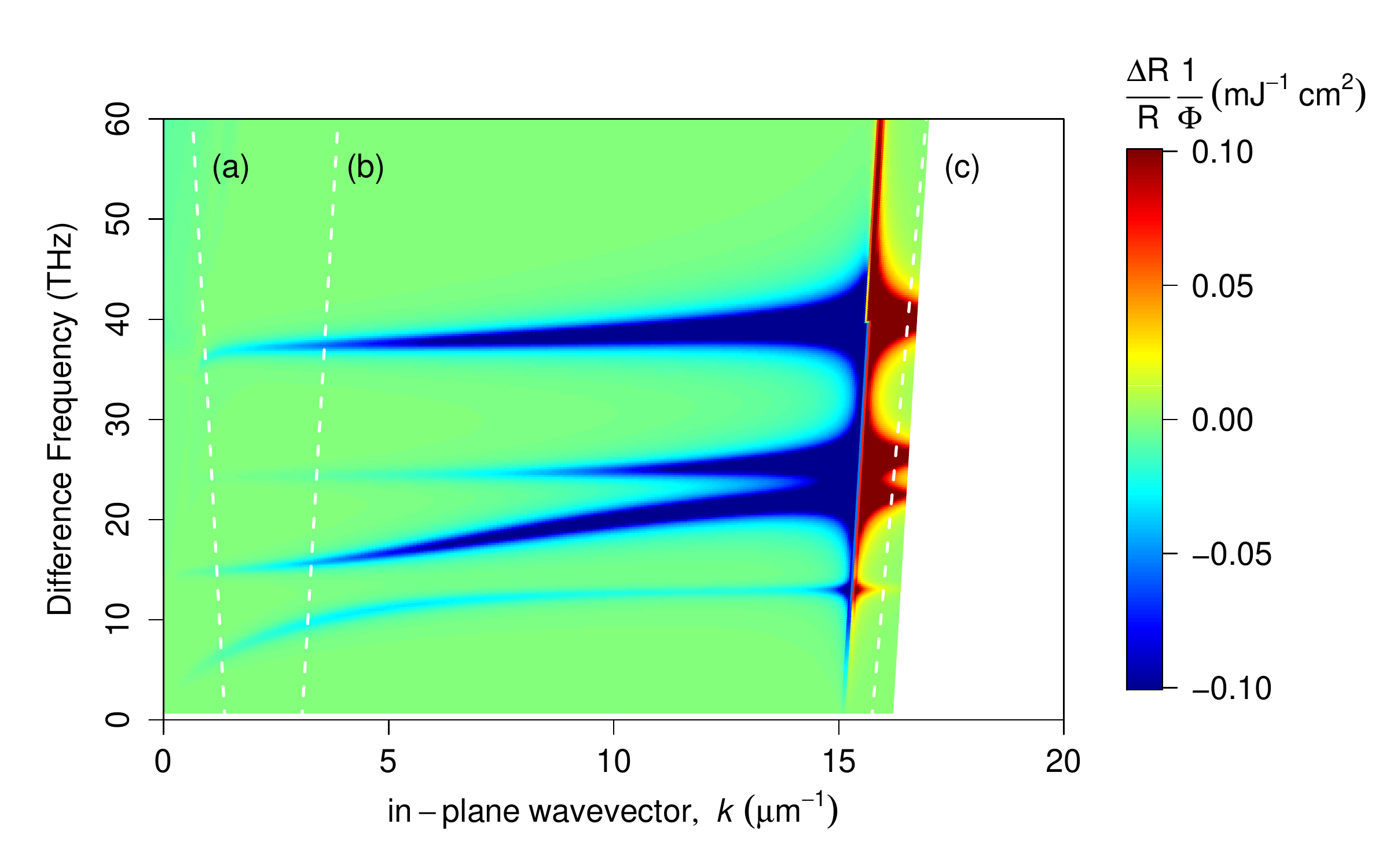}
\end{center}
\caption{The numerical solution for the differential probe reflectance normalized by fluence, calculated using the model outlined in the SI. The white dotted lines indicate the region of the dispersion relation probed by the experimental geometries shown in fig.~\ref{fig:dispersion_compilation}.\label{fig:theory_dispersion}}
\end{figure}
In Fig.~\ref{fig:theory_dispersion}, we plot the numerical solution for the differential probe reflectance, normalized by fluence, for the simplified case of continuous plane-wave pump and probe beams. While this simple model ignores the non-equilibrium nature of the excitation, as observed in experiment, we show below that it is sufficient to describe some of the salient features of our results. Similar to Fig.~\ref{fig:dispersion_compilation}, the differential reflectance is plotted versus difference frequency and in-plane wavevector, whose values are scanned by continuously varying the probe incidence angle and pump wavelength. The probe wavelength and pump angles are fixed at $615~\mathrm{nm}$ and $50^\circ$, respectively, and the pump and probe intensities are chosen to be 10 and $0.1~\mathrm{W/\mu m^2}$, to closely correspond to the configuration in Fig.~\ref{fig:dispersion_compilation}(b). It can be seen that the simulation qualitatively produces the main features of Fig.~\ref{fig:dispersion_compilation}. In particular, the change in the sign of differential reflectance at the Brewster angle is clearly observed, as is the enhancement of the signal when the difference frequency and wavevector align with the plasmon dispersion relation.

The model also reproduces some of the main features arising from different coupling efficiencies to different bands (fig.~\ref{fig:theory_dispersion}). Generally, the highest coupling efficiency occurs for the dispersion regions which are most `plasmon-like' in origin (see additionally fig.~\ref{fig:theory_eta}).  This is most obvious comparing the data in fig.~\ref{fig:dispersion_compilation}(c) and fig.~\ref{fig:theory_dispersion}(c), where the coupling to the upper bands is much stronger than the lowest band in both model and experiment. For lower wavevector cases, the coupling to the highest band is overestimated in the model compared to the experiment. 	This could possibly be caused by frequency-dependant losses in the graphene sheet unaccounted for in our simple model. 
The model additionally reproduces the increasing absolute signal strength with increasing wavevector observed in experiment, which is a consequence of both larger changes in the reflection coefficient for a corresponding change in absorption for higher angles, and due to spatial dispersion\cite{Mikhailov2011} in the signal. Indeed, it can be shown that the magnitude of enhancement is proportional to the square of the plasmon quality factor, $Q^2$ (see SI), in agreement with predictions from ref.~\cite{Yao2014}. 

In addition to the surface plasmon resonance conditions, for the highest wavevector region in fig.~\ref{fig:dispersion_compilation}(c) there is an additional resonant enhancement found at low frequencies~$< 3~\mathrm{THz}$ measured in experiment that is not reproduced in our model (fig.~\ref{fig:theory_dispersion}(c)). The position of this peak lies within the expected region of intraband transitions in graphene, indicated by the dotted line in fig.~\ref{fig:dispersion_compilation}. This feature is also largely polarization independent, unlike the enhancements we attribute to surface plasmon coupling (see SI, fig.~\ref{fig:polarisation}).

In principle, Eq.~\ref{eq:tprobemain} can be inverted to allow an experimental determination of the nonlinear conductivity $\sigma^{(2)}$, given transmission or reflection data. This is difficult in the present setup, in part given the broad bandwidth of the pulses, uncertainty over some system parameters, and difficulty of investigating a large number of angles to quantify possible wavevector and frequency dependence of $\sigma^{(2)}$. However, as an estimate, we take the simplest possible model, in which the effective nonlinear susceptibility $\chi^{(2)}$ is frequency and wavevector-independent. This corresponds to a nonlinear conductivity function obeying $\sigma^{(2)}(\omega)=i |\sigma^{(2)} (\omega_{probe} )|(\omega / \omega_{probe} )$, where the value at the (fixed) probe frequency represents a single fitting parameter. We find that a value of $|\sigma^{(2)} (\omega_{probe})|\approx 2.4\times 10^{-12}~\mathrm{A\cdot m/V^2}$ produces the same peak signal as observed in Fig. 3(b). It should be emphasized that this represents a rather conservative estimate of $\sigma^{(2)} (\omega)$. In particular, the mobility of $\mu\approx 2000$~cm${}^2$/Vs corresponds to a plasmon linewidth of $\gamma=2\pi\times1.6$~THz that is narrower than the measurement bandwidth, indicating that only a fraction of the pulse can efficiently excite plasmons. Reducing measurement bandwidth could therefore give rise to greater coupling to the surface plasmons, while also reducing the effects of non-equilibrium carriers on the measurements. While a comparison to a bulk nonlinear crystal is not directly meaningful, it is nonetheless interesting to note that a bulk nonlinear crystal with the thickness $t \approx 0.3 ~\mathrm{nm}$ of a graphene layer would require a nonlinear susceptibility of $\chi^{(2)}\sim|\sigma^{(2)} (\omega_{probe} )|/(\varepsilon_0 \omega_{probe} t)\sim 3\times 10^{-7}~\mathrm{m/V}$ to produce the equivalent in-plane nonlinear currents. This value is approximately 3 orders of magnitude larger than in GaAs.

Finally, from the inferred value of $\sigma^{(2)}$ and the input beam parameters, our model enables us to estimate the conversion efficiency $\eta$ of pump photons to plasmons (see SI). We find a value of $\eta \approx 6\times 10^{-6}$, while noting that the actual conversion could be significantly higher with narrow pulses, again as the estimated value of $\sigma^{(2)}$ does not account for the large pulse bandwidth. We note that this experimentally obtained value of $\eta$ is of the same order as predicted in ref.~\cite{Yao2014}, once adjusted for our experimental parameters.

In summary, surface plasmon generation in a planar single layer graphene crystal has been demonstrated using nonlinear wave mixing of visible frequency light. We observe enhancements in the nonlinear, differential reflection of light for various frequencies and phase-matching conditions which agree well with the surface plasmon-phonon dispersion for graphene on quartz substrates. In contrast to near-field infrared scattering\cite{Fei2012,Chen2012}, our approach may be used to access an distinctly broad frequency range, including the ordinarily hard to reach THz “gap”\cite{Tonouchi2007}. Moreover, by careful manipulating the phase-matching conditions, we show that one can generate surface plasmons with a defined wavevector, with an efficiency approaching $10^{-5}$. This efficiency by no means represents a fundamental limit, and we believe that it could in principle be pushed towards a $10^{-2}$ level with future adjustments, such as increasing the surface plasmon Q factor from $\sim 5$ to $\sim30$ with lattice-matched hBN substrates\cite{Woessner2014a}, equalizing the intensities of the pump and probe beams (see SI, fig. \ref{fig:low_fluence}), or the use of narrower bandwidth pulses. Moreover, in principle, our approach could be extended to higher or lower frequencies, regions that are generally hard to access using current approaches\cite{Abajo2014}. 

\section{Methods}

\subsection{Sample Preparation}
Samples for our experiments are fabricated from commercially grown CVD graphene on copper foil (graphene supermarket). Transfer to quartz substrates was performed in house via a standard metal etching and float technique using Ammonium persulfate to etch the copper and PMMA as a support structure. Combined resistance and Raman spectroscopy\cite{Lee2012} give an estimated mobility of the samples of around $2000~\mathrm{cm^2/Vs}$ and a natural Fermi energy of $\sim 300~\mathrm{meV}$. Raman imaging indicates that the graphene is nominally single layer, with $\sim 80\%$ coverage of the substrate.

\section{Acknowledgements}
This research has been supported by the European Commission (FP7-ICT-2013-613024-GRASP) and EPSRC fellowship (EP/K041215/1). DEC also acknowledges support from Fundacio Privada Cellex Barcelona and the Ramon y Cajal program. The authors would like to thank Javier Garc\'ia de Abajo and Marinko Jablan for useful discussions, and Nick Cole for technical assistance.

\section{Author Contributions}

\section{Competing Financial Interests}
The authors declare no competing financial interests.

\section{Figure Captions}


\begin{thebibliography}{38}
\expandafter\ifx\csname natexlab\endcsname\relax\def\natexlab#1{#1}\fi
\expandafter\ifx\csname bibnamefont\endcsname\relax
  \def\bibnamefont#1{#1}\fi
\expandafter\ifx\csname bibfnamefont\endcsname\relax
  \def\bibfnamefont#1{#1}\fi
\expandafter\ifx\csname citenamefont\endcsname\relax
  \def\citenamefont#1{#1}\fi
\expandafter\ifx\csname url\endcsname\relax
  \def\url#1{\texttt{#1}}\fi
\expandafter\ifx\csname urlprefix\endcsname\relax\def\urlprefix{URL }\fi
\providecommand{\bibinfo}[2]{#2}
\providecommand{\eprint}[2][]{\url{#2}}

\bibitem[{\citenamefont{Grigorenko et~al.}(2012)\citenamefont{Grigorenko,
  Polini, and Novoselov}}]{Grigorenko2012}
\bibinfo{author}{\bibfnamefont{A.~N.} \bibnamefont{Grigorenko}},
  \bibinfo{author}{\bibfnamefont{M.}~\bibnamefont{Polini}}, \bibnamefont{and}
  \bibinfo{author}{\bibfnamefont{K.~S.} \bibnamefont{Novoselov}},
  \bibinfo{journal}{Nat. Photonics} \textbf{\bibinfo{volume}{6}},
  \bibinfo{pages}{749 } (\bibinfo{year}{2012}), ISSN \bibinfo{issn}{1749-4885},
  \eprint{1301.4241},
  \urlprefix\url{http://dx.doi.org/10.1038/nphoton.2012.262}.

\bibitem[{\citenamefont{de~Abajo}(2014)}]{Abajo2014}
\bibinfo{author}{\bibfnamefont{F.~G.} \bibnamefont{de~Abajo}},
  \bibinfo{journal}{ACS Photonics} \textbf{\bibinfo{volume}{1}},
  \bibinfo{pages}{135} (\bibinfo{year}{2014}), \eprint{arXiv:1402.1969v2},
  \urlprefix\url{http://pubs.acs.org/doi/abs/10.1021/ph400147y}.

\bibitem[{\citenamefont{Koppens et~al.}(2011)\citenamefont{Koppens, Chang, and
  de~Abajo}}]{koppensnl}
\bibinfo{author}{\bibfnamefont{F.~H.~L.} \bibnamefont{Koppens}},
  \bibinfo{author}{\bibfnamefont{D.~E.} \bibnamefont{Chang}}, \bibnamefont{and}
  \bibinfo{author}{\bibfnamefont{F.~J.} \bibnamefont{de~Abajo}},
  \bibinfo{journal}{Nano Letters} \textbf{\bibinfo{volume}{11}},
  \bibinfo{pages}{3370} (\bibinfo{year}{2011}),
  \urlprefix\url{http://dx.doi.org/10.1021/nl201771h}.

\bibitem[{\citenamefont{Craciun et~al.}(2011)\citenamefont{Craciun, Russo,
  Yamamoto, and Tarucha}}]{Craciun2011}
\bibinfo{author}{\bibfnamefont{M.~F.} \bibnamefont{Craciun}},
  \bibinfo{author}{\bibfnamefont{S.}~\bibnamefont{Russo}},
  \bibinfo{author}{\bibfnamefont{M.}~\bibnamefont{Yamamoto}}, \bibnamefont{and}
  \bibinfo{author}{\bibfnamefont{S.}~\bibnamefont{Tarucha}},
  \bibinfo{journal}{Nano Today} \textbf{\bibinfo{volume}{6}},
  \bibinfo{pages}{42} (\bibinfo{year}{2011}), ISSN \bibinfo{issn}{17480132},
  \urlprefix\url{http://linkinghub.elsevier.com/retrieve/pii/S1748013210001623}.

\bibitem[{\citenamefont{Hendry et~al.}(2010)\citenamefont{Hendry, Hale, Moger,
  Savchenko, and Mikhailov}}]{Hendry2010a}
\bibinfo{author}{\bibfnamefont{E.}~\bibnamefont{Hendry}},
  \bibinfo{author}{\bibfnamefont{P.~J.} \bibnamefont{Hale}},
  \bibinfo{author}{\bibfnamefont{J.}~\bibnamefont{Moger}},
  \bibinfo{author}{\bibfnamefont{A.~K.} \bibnamefont{Savchenko}},
  \bibnamefont{and} \bibinfo{author}{\bibfnamefont{S.~A.}
  \bibnamefont{Mikhailov}}, \bibinfo{journal}{Physical Review Letters}
  \textbf{\bibinfo{volume}{105}}, \bibinfo{pages}{097401}
  (\bibinfo{year}{2010}), ISSN \bibinfo{issn}{0031-9007},
  \urlprefix\url{http://link.aps.org/doi/10.1103/PhysRevLett.105.097401}.

\bibitem[{\citenamefont{Brar et~al.}(2013)\citenamefont{Brar, Jang, Sherrott,
  Lopez, and Atwater}}]{Brar2013}
\bibinfo{author}{\bibfnamefont{V.~W.} \bibnamefont{Brar}},
  \bibinfo{author}{\bibfnamefont{M.~S.} \bibnamefont{Jang}},
  \bibinfo{author}{\bibfnamefont{M.}~\bibnamefont{Sherrott}},
  \bibinfo{author}{\bibfnamefont{J.~J.} \bibnamefont{Lopez}}, \bibnamefont{and}
  \bibinfo{author}{\bibfnamefont{H.~a.} \bibnamefont{Atwater}},
  \bibinfo{journal}{Nano Letters} \textbf{\bibinfo{volume}{13}},
  \bibinfo{pages}{2541} (\bibinfo{year}{2013}), ISSN \bibinfo{issn}{15306984}.

\bibitem[{\citenamefont{Fei et~al.}(2012)\citenamefont{Fei, Rodin, Andreev,
  Bao, McLeod, Wagner, Zhang, Zhao, Thiemens, Dominguez et~al.}}]{Fei2012}
\bibinfo{author}{\bibfnamefont{Z.}~\bibnamefont{Fei}},
  \bibinfo{author}{\bibfnamefont{A.~S.} \bibnamefont{Rodin}},
  \bibinfo{author}{\bibfnamefont{G.~O.} \bibnamefont{Andreev}},
  \bibinfo{author}{\bibfnamefont{W.}~\bibnamefont{Bao}},
  \bibinfo{author}{\bibfnamefont{A.~S.} \bibnamefont{McLeod}},
  \bibinfo{author}{\bibfnamefont{M.}~\bibnamefont{Wagner}},
  \bibinfo{author}{\bibfnamefont{L.~M.} \bibnamefont{Zhang}},
  \bibinfo{author}{\bibfnamefont{Z.}~\bibnamefont{Zhao}},
  \bibinfo{author}{\bibfnamefont{M.}~\bibnamefont{Thiemens}},
  \bibinfo{author}{\bibfnamefont{G.}~\bibnamefont{Dominguez}},
  \bibnamefont{et~al.}, \bibinfo{journal}{Nature}
  \textbf{\bibinfo{volume}{487}}, \bibinfo{pages}{82} (\bibinfo{year}{2012}),
  ISSN \bibinfo{issn}{1476-4687},
  \urlprefix\url{http://www.ncbi.nlm.nih.gov/pubmed/22722866}.

\bibitem[{\citenamefont{Chen et~al.}(2012)\citenamefont{Chen, Badioli,
  Alonso-Gonz\'{a}lez, Thongrattanasiri, Huth, Osmond, Spasenovi\'{c}, Centeno,
  Pesquera, Godignon et~al.}}]{Chen2012}
\bibinfo{author}{\bibfnamefont{J.}~\bibnamefont{Chen}},
  \bibinfo{author}{\bibfnamefont{M.}~\bibnamefont{Badioli}},
  \bibinfo{author}{\bibfnamefont{P.}~\bibnamefont{Alonso-Gonz\'{a}lez}},
  \bibinfo{author}{\bibfnamefont{S.}~\bibnamefont{Thongrattanasiri}},
  \bibinfo{author}{\bibfnamefont{F.}~\bibnamefont{Huth}},
  \bibinfo{author}{\bibfnamefont{J.}~\bibnamefont{Osmond}},
  \bibinfo{author}{\bibfnamefont{M.}~\bibnamefont{Spasenovi\'{c}}},
  \bibinfo{author}{\bibfnamefont{A.}~\bibnamefont{Centeno}},
  \bibinfo{author}{\bibfnamefont{A.}~\bibnamefont{Pesquera}},
  \bibinfo{author}{\bibfnamefont{P.}~\bibnamefont{Godignon}},
  \bibnamefont{et~al.}, \bibinfo{journal}{Nature}
  \textbf{\bibinfo{volume}{487}}, \bibinfo{pages}{77} (\bibinfo{year}{2012}),
  ISSN \bibinfo{issn}{1476-4687},
  \urlprefix\url{http://www.ncbi.nlm.nih.gov/pubmed/22722861}.

\bibitem[{\citenamefont{Bludov et~al.}(2013)\citenamefont{Bludov, Ferreira,
  Peres, and Vasilevskiy}}]{Bludov2013}
\bibinfo{author}{\bibfnamefont{Y.~V.} \bibnamefont{Bludov}},
  \bibinfo{author}{\bibfnamefont{A.}~\bibnamefont{Ferreira}},
  \bibinfo{author}{\bibfnamefont{N.~M.~R.} \bibnamefont{Peres}},
  \bibnamefont{and} \bibinfo{author}{\bibfnamefont{M.~I.}
  \bibnamefont{Vasilevskiy}}, \bibinfo{journal}{International Journal of Modern
  Physics B} \textbf{\bibinfo{volume}{27}}, \bibinfo{pages}{1341001}
  (\bibinfo{year}{2013}), ISSN \bibinfo{issn}{0217-9792},
  \urlprefix\url{http://www.worldscientific.com/doi/abs/10.1142/S0217979213410014}.

\bibitem[{\citenamefont{Gao et~al.}(2012)\citenamefont{Gao, Shu, Qiu, and
  Xu}}]{Gao2012a}
\bibinfo{author}{\bibfnamefont{W.}~\bibnamefont{Gao}},
  \bibinfo{author}{\bibfnamefont{J.}~\bibnamefont{Shu}},
  \bibinfo{author}{\bibfnamefont{C.}~\bibnamefont{Qiu}}, \bibnamefont{and}
  \bibinfo{author}{\bibfnamefont{Q.}~\bibnamefont{Xu}}, \bibinfo{journal}{ACS
  Nano} \textbf{\bibinfo{volume}{6}}, \bibinfo{pages}{7806}
  (\bibinfo{year}{2012}), ISSN \bibinfo{issn}{19360851}.

\bibitem[{\citenamefont{Alonso-Gonzalez
  et~al.}(2014)\citenamefont{Alonso-Gonzalez, Nikitin, Golmar, Centeno,
  Pesquera, Velez, Chen, Navickaite, Koppens, Zurutuza
  et~al.}}]{Alonso-Gonzalez2014}
\bibinfo{author}{\bibfnamefont{P.}~\bibnamefont{Alonso-Gonzalez}},
  \bibinfo{author}{\bibfnamefont{A.~Y.} \bibnamefont{Nikitin}},
  \bibinfo{author}{\bibfnamefont{F.}~\bibnamefont{Golmar}},
  \bibinfo{author}{\bibfnamefont{A.}~\bibnamefont{Centeno}},
  \bibinfo{author}{\bibfnamefont{A.}~\bibnamefont{Pesquera}},
  \bibinfo{author}{\bibfnamefont{S.}~\bibnamefont{Velez}},
  \bibinfo{author}{\bibfnamefont{J.}~\bibnamefont{Chen}},
  \bibinfo{author}{\bibfnamefont{G.}~\bibnamefont{Navickaite}},
  \bibinfo{author}{\bibfnamefont{F.}~\bibnamefont{Koppens}},
  \bibinfo{author}{\bibfnamefont{A.}~\bibnamefont{Zurutuza}},
  \bibnamefont{et~al.}, \bibinfo{journal}{Science}
  \textbf{\bibinfo{volume}{344}}, \bibinfo{pages}{1369} (\bibinfo{year}{2014}),
  ISSN \bibinfo{issn}{0036-8075},
  \urlprefix\url{http://www.sciencemag.org/cgi/doi/10.1126/science.1253202}.

\bibitem[{\citenamefont{Gao et~al.}(2013)\citenamefont{Gao, Shi, Jin, Shu,
  Zhang, Vajtai, Ajayan, Kono, and Xu}}]{Gao2013}
\bibinfo{author}{\bibfnamefont{W.}~\bibnamefont{Gao}},
  \bibinfo{author}{\bibfnamefont{G.}~\bibnamefont{Shi}},
  \bibinfo{author}{\bibfnamefont{Z.}~\bibnamefont{Jin}},
  \bibinfo{author}{\bibfnamefont{J.}~\bibnamefont{Shu}},
  \bibinfo{author}{\bibfnamefont{Q.}~\bibnamefont{Zhang}},
  \bibinfo{author}{\bibfnamefont{R.}~\bibnamefont{Vajtai}},
  \bibinfo{author}{\bibfnamefont{P.~M.} \bibnamefont{Ajayan}},
  \bibinfo{author}{\bibfnamefont{J.}~\bibnamefont{Kono}}, \bibnamefont{and}
  \bibinfo{author}{\bibfnamefont{Q.}~\bibnamefont{Xu}}, \bibinfo{journal}{Nano
  letters} \textbf{\bibinfo{volume}{13}}, \bibinfo{pages}{3698}
  (\bibinfo{year}{2013}), ISSN \bibinfo{issn}{1530-6992},
  \urlprefix\url{http://www.ncbi.nlm.nih.gov/pubmed/23895501}.

\bibitem[{\citenamefont{Farhat et~al.}(2013)\citenamefont{Farhat, Guenneau, and
  Bağcı}}]{Farhat2013}
\bibinfo{author}{\bibfnamefont{M.}~\bibnamefont{Farhat}},
  \bibinfo{author}{\bibfnamefont{S.}~\bibnamefont{Guenneau}}, \bibnamefont{and}
  \bibinfo{author}{\bibfnamefont{H.}~\bibnamefont{Bağcı}},
  \bibinfo{journal}{Physical Review Letters} \textbf{\bibinfo{volume}{111}},
  \bibinfo{pages}{237404} (\bibinfo{year}{2013}), ISSN
  \bibinfo{issn}{0031-9007},
  \urlprefix\url{http://link.aps.org/doi/10.1103/PhysRevLett.111.237404}.

\bibitem[{\citenamefont{Schiefele et~al.}(2013)\citenamefont{Schiefele,
  Pedr\'{o}s, Sols, Calle, and Guinea}}]{Schiefele2013}
\bibinfo{author}{\bibfnamefont{J.}~\bibnamefont{Schiefele}},
  \bibinfo{author}{\bibfnamefont{J.}~\bibnamefont{Pedr\'{o}s}},
  \bibinfo{author}{\bibfnamefont{F.}~\bibnamefont{Sols}},
  \bibinfo{author}{\bibfnamefont{F.}~\bibnamefont{Calle}}, \bibnamefont{and}
  \bibinfo{author}{\bibfnamefont{F.}~\bibnamefont{Guinea}},
  \bibinfo{journal}{Physical Review Letters} \textbf{\bibinfo{volume}{111}},
  \bibinfo{pages}{237405} (\bibinfo{year}{2013}), ISSN
  \bibinfo{issn}{0031-9007}, \eprint{arXiv:1309.0767v1},
  \urlprefix\url{http://link.aps.org/doi/10.1103/PhysRevLett.111.237405}.

\bibitem[{\citenamefont{Manzoni et~al.}(2014)\citenamefont{Manzoni, Garc,
  Abajo, and Chang}}]{Manzoni}
\bibinfo{author}{\bibfnamefont{M.~T.} \bibnamefont{Manzoni}},
  \bibinfo{author}{\bibfnamefont{F.~J.} \bibnamefont{Garc}},
  \bibinfo{author}{\bibfnamefont{D.}~\bibnamefont{Abajo}}, \bibnamefont{and}
  \bibinfo{author}{\bibfnamefont{D.~E.} \bibnamefont{Chang}},
  \bibinfo{journal}{Preprint at http://arxiv.org/abs/1406.4360}
  (\bibinfo{year}{2014}), \eprint{1406.4360v1}.

\bibitem[{\citenamefont{Dean and van Driel}(2010)}]{Dean2010}
\bibinfo{author}{\bibfnamefont{J.~J.} \bibnamefont{Dean}} \bibnamefont{and}
  \bibinfo{author}{\bibfnamefont{H.~M.} \bibnamefont{van Driel}},
  \bibinfo{journal}{Physical Review B} \textbf{\bibinfo{volume}{82}},
  \bibinfo{pages}{125411} (\bibinfo{year}{2010}), ISSN
  \bibinfo{issn}{1098-0121},
  \urlprefix\url{http://link.aps.org/doi/10.1103/PhysRevB.82.125411}.

\bibitem[{\citenamefont{Hong et~al.}(2013)\citenamefont{Hong, Dadap, Petrone,
  Yeh, Hone, and Osgood}}]{Hong2013}
\bibinfo{author}{\bibfnamefont{S.-Y.} \bibnamefont{Hong}},
  \bibinfo{author}{\bibfnamefont{J.~I.} \bibnamefont{Dadap}},
  \bibinfo{author}{\bibfnamefont{N.}~\bibnamefont{Petrone}},
  \bibinfo{author}{\bibfnamefont{P.-C.} \bibnamefont{Yeh}},
  \bibinfo{author}{\bibfnamefont{J.}~\bibnamefont{Hone}}, \bibnamefont{and}
  \bibinfo{author}{\bibfnamefont{R.~M.} \bibnamefont{Osgood}},
  \bibinfo{journal}{Physical Review X} \textbf{\bibinfo{volume}{3}},
  \bibinfo{pages}{021014} (\bibinfo{year}{2013}), ISSN
  \bibinfo{issn}{2160-3308},
  \urlprefix\url{http://link.aps.org/doi/10.1103/PhysRevX.3.021014}.

\bibitem[{\citenamefont{Gu et~al.}(2012)\citenamefont{Gu, Petrone, McMillan,
  van~der Zande, Yu, Lo, Kwong, Hone, and Wong}}]{Gu2012}
\bibinfo{author}{\bibfnamefont{T.}~\bibnamefont{Gu}},
  \bibinfo{author}{\bibfnamefont{N.}~\bibnamefont{Petrone}},
  \bibinfo{author}{\bibfnamefont{J.~F.} \bibnamefont{McMillan}},
  \bibinfo{author}{\bibfnamefont{A.}~\bibnamefont{van~der Zande}},
  \bibinfo{author}{\bibfnamefont{M.}~\bibnamefont{Yu}},
  \bibinfo{author}{\bibfnamefont{G.~Q.} \bibnamefont{Lo}},
  \bibinfo{author}{\bibfnamefont{D.~L.} \bibnamefont{Kwong}},
  \bibinfo{author}{\bibfnamefont{J.}~\bibnamefont{Hone}}, \bibnamefont{and}
  \bibinfo{author}{\bibfnamefont{C.~W.} \bibnamefont{Wong}},
  \bibinfo{journal}{Nature Photonics} \textbf{\bibinfo{volume}{6}},
  \bibinfo{pages}{554} (\bibinfo{year}{2012}), ISSN \bibinfo{issn}{1749-4885},
  \eprint{1205.4333},
  \urlprefix\url{http://www.nature.com/doifinder/10.1038/nphoton.2012.147}.

\bibitem[{\citenamefont{Gullans et~al.}(2013)\citenamefont{Gullans, Chang,
  Koppens, de~Abajo, and Lukin}}]{Gullans2013}
\bibinfo{author}{\bibfnamefont{M.}~\bibnamefont{Gullans}},
  \bibinfo{author}{\bibfnamefont{D.~E.} \bibnamefont{Chang}},
  \bibinfo{author}{\bibfnamefont{F.~H.~L.} \bibnamefont{Koppens}},
  \bibinfo{author}{\bibfnamefont{F.~J.~G.} \bibnamefont{de~Abajo}},
  \bibnamefont{and} \bibinfo{author}{\bibfnamefont{M.~D.} \bibnamefont{Lukin}},
  \bibinfo{journal}{Physical Review Letters} \textbf{\bibinfo{volume}{111}},
  \bibinfo{pages}{1} (\bibinfo{year}{2013}), ISSN \bibinfo{issn}{00319007},
  \eprint{arXiv:1309.2651v1}.

\bibitem[{\citenamefont{Mikhailov}(2011)}]{Mikhailov2011}
\bibinfo{author}{\bibfnamefont{S.~A.} \bibnamefont{Mikhailov}},
  \bibinfo{journal}{Phys. Rev. B} \textbf{\bibinfo{volume}{84}},
  \bibinfo{pages}{045432} (\bibinfo{year}{2011}), ISSN
  \bibinfo{issn}{10980121}.

\bibitem[{\citenamefont{Renger et~al.}(2009)\citenamefont{Renger, Quidant, van
  Hulst, Palomba, and Novotny}}]{Renger2009}
\bibinfo{author}{\bibfnamefont{J.}~\bibnamefont{Renger}},
  \bibinfo{author}{\bibfnamefont{R.}~\bibnamefont{Quidant}},
  \bibinfo{author}{\bibfnamefont{N.}~\bibnamefont{van Hulst}},
  \bibinfo{author}{\bibfnamefont{S.}~\bibnamefont{Palomba}}, \bibnamefont{and}
  \bibinfo{author}{\bibfnamefont{L.}~\bibnamefont{Novotny}},
  \bibinfo{journal}{Physical Review Letters} \textbf{\bibinfo{volume}{103}},
  \bibinfo{pages}{266802} (\bibinfo{year}{2009}), ISSN
  \bibinfo{issn}{0031-9007},
  \urlprefix\url{http://link.aps.org/doi/10.1103/PhysRevLett.103.266802}.

\bibitem[{\citenamefont{Palomba and Novotny}(2008)}]{Palomba2008}
\bibinfo{author}{\bibfnamefont{S.}~\bibnamefont{Palomba}} \bibnamefont{and}
  \bibinfo{author}{\bibfnamefont{L.}~\bibnamefont{Novotny}},
  \bibinfo{journal}{Physical Review Letters} \textbf{\bibinfo{volume}{101}},
  \bibinfo{pages}{056802} (\bibinfo{year}{2008}), ISSN
  \bibinfo{issn}{0031-9007},
  \urlprefix\url{http://link.aps.org/doi/10.1103/PhysRevLett.101.056802}.

\bibitem[{\citenamefont{Yao et~al.}(2014)\citenamefont{Yao, Tokman, and
  Belyanin}}]{Yao2014}
\bibinfo{author}{\bibfnamefont{X.}~\bibnamefont{Yao}},
  \bibinfo{author}{\bibfnamefont{M.}~\bibnamefont{Tokman}}, \bibnamefont{and}
  \bibinfo{author}{\bibfnamefont{A.}~\bibnamefont{Belyanin}},
  \bibinfo{journal}{Physical Review Letters} \textbf{\bibinfo{volume}{112}},
  \bibinfo{pages}{055501} (\bibinfo{year}{2014}), ISSN
  \bibinfo{issn}{00319007}, \eprint{arXiv:1308.2005v1},
  \urlprefix\url{http://arxiv.org/abs/1308.2005
  http://link.aps.org/doi/10.1103/PhysRevLett.112.055501}.

\bibitem[{\citenamefont{Sun et~al.}(2014)\citenamefont{Sun, Qiao, and
  Sun}}]{Sun2014}
\bibinfo{author}{\bibfnamefont{Y.}~\bibnamefont{Sun}},
  \bibinfo{author}{\bibfnamefont{G.}~\bibnamefont{Qiao}}, \bibnamefont{and}
  \bibinfo{author}{\bibfnamefont{G.}~\bibnamefont{Sun}},
  \bibinfo{journal}{Optics Express} \textbf{\bibinfo{volume}{22}},
  \bibinfo{pages}{27880} (\bibinfo{year}{2014}), ISSN
  \bibinfo{issn}{1094-4087}, \eprint{arXiv:1405.3320},
  \urlprefix\url{http://www.opticsinfobase.org/abstract.cfm?URI=oe-22-23-27880}.

\bibitem[{\citenamefont{Boyd}(2008)}]{Boyd2008}
\bibinfo{author}{\bibfnamefont{R.~W.} \bibnamefont{Boyd}},
  \emph{\bibinfo{title}{{Nonlinear Optics}}} (\bibinfo{publisher}{Elsevier},
  \bibinfo{year}{2008}), \bibinfo{edition}{third edit} ed., ISBN
  \bibinfo{isbn}{978-0-12-369470-6}.

\bibitem[{\citenamefont{Alexeev et~al.}(2013)\citenamefont{Alexeev, Moger, and
  Hendry}}]{Alexeev2013}
\bibinfo{author}{\bibfnamefont{E.}~\bibnamefont{Alexeev}},
  \bibinfo{author}{\bibfnamefont{J.}~\bibnamefont{Moger}}, \bibnamefont{and}
  \bibinfo{author}{\bibfnamefont{E.}~\bibnamefont{Hendry}},
  \bibinfo{journal}{Applied Physics Letters} \textbf{\bibinfo{volume}{103}},
  \bibinfo{pages}{2011} (\bibinfo{year}{2013}), ISSN \bibinfo{issn}{00036951},
  \eprint{1310.0970}.

\bibitem[{\citenamefont{Brida et~al.}(2013)\citenamefont{Brida, Tomadin,
  Manzoni, Kim, Lombardo, Milana, Nair, Novoselov, Ferrari, Cerullo
  et~al.}}]{Brida2013}
\bibinfo{author}{\bibfnamefont{D.}~\bibnamefont{Brida}},
  \bibinfo{author}{\bibfnamefont{A.}~\bibnamefont{Tomadin}},
  \bibinfo{author}{\bibfnamefont{C.}~\bibnamefont{Manzoni}},
  \bibinfo{author}{\bibfnamefont{Y.~J.} \bibnamefont{Kim}},
  \bibinfo{author}{\bibfnamefont{A.}~\bibnamefont{Lombardo}},
  \bibinfo{author}{\bibfnamefont{S.}~\bibnamefont{Milana}},
  \bibinfo{author}{\bibfnamefont{R.~R.} \bibnamefont{Nair}},
  \bibinfo{author}{\bibfnamefont{K.~S.} \bibnamefont{Novoselov}},
  \bibinfo{author}{\bibfnamefont{A.~C.} \bibnamefont{Ferrari}},
  \bibinfo{author}{\bibfnamefont{G.}~\bibnamefont{Cerullo}},
  \bibnamefont{et~al.}, \bibinfo{journal}{Nature communications}
  \textbf{\bibinfo{volume}{4}}, \bibinfo{pages}{1987} (\bibinfo{year}{2013}),
  ISSN \bibinfo{issn}{2041-1723},
  \urlprefix\url{http://www.ncbi.nlm.nih.gov/pubmed/23770933}.

\bibitem[{\citenamefont{Dawlaty et~al.}(2008)\citenamefont{Dawlaty, Shivaraman,
  Chandrashekhar, Rana, and Spencer}}]{Dawlaty2008}
\bibinfo{author}{\bibfnamefont{J.~M.} \bibnamefont{Dawlaty}},
  \bibinfo{author}{\bibfnamefont{S.}~\bibnamefont{Shivaraman}},
  \bibinfo{author}{\bibfnamefont{M.}~\bibnamefont{Chandrashekhar}},
  \bibinfo{author}{\bibfnamefont{F.}~\bibnamefont{Rana}}, \bibnamefont{and}
  \bibinfo{author}{\bibfnamefont{M.~G.} \bibnamefont{Spencer}},
  \bibinfo{journal}{Applied Physics Letters} \textbf{\bibinfo{volume}{92}},
  \bibinfo{pages}{21} (\bibinfo{year}{2008}), ISSN \bibinfo{issn}{00036951}.
  
  \bibitem[{\citenamefont{Hale et~al.}(2011)\citenamefont{Hale, Hornett, Moger,
  Horsell, and Hendry}}]{Hale2011b}
\bibinfo{author}{\bibfnamefont{P.~J.} \bibnamefont{Hale}},
  \bibinfo{author}{\bibfnamefont{S.~M.} \bibnamefont{Hornett}},
  \bibinfo{author}{\bibfnamefont{J.}~\bibnamefont{Moger}},
  \bibinfo{author}{\bibfnamefont{D.~W.} \bibnamefont{Horsell}},
  \bibnamefont{and} \bibinfo{author}{\bibfnamefont{E.}~\bibnamefont{Hendry}},
  \bibinfo{journal}{Phys. Rev. B} \textbf{\bibinfo{volume}{83}},
  \bibinfo{pages}{121404} (\bibinfo{year}{2011}), ISSN
  \bibinfo{issn}{10980121}, \eprint{1012.3927}.


\bibitem[{\citenamefont{Borri et~al.}(2000)\citenamefont{Borri, Romstad,
  Langbein, Kelly, Mork, and Hvam}}]{Borri2000}
\bibinfo{author}{\bibfnamefont{P.}~\bibnamefont{Borri}},
  \bibinfo{author}{\bibfnamefont{F.}~\bibnamefont{Romstad}},
  \bibinfo{author}{\bibfnamefont{W.}~\bibnamefont{Langbein}},
  \bibinfo{author}{\bibfnamefont{a.}~\bibnamefont{Kelly}},
  \bibinfo{author}{\bibfnamefont{J.}~\bibnamefont{Mork}}, \bibnamefont{and}
  \bibinfo{author}{\bibfnamefont{J.}~\bibnamefont{Hvam}},
  \bibinfo{journal}{Optics express} \textbf{\bibinfo{volume}{7}},
  \bibinfo{pages}{107} (\bibinfo{year}{2000}), ISSN \bibinfo{issn}{1094-4087}.

\bibitem[{\citenamefont{Luxmoore et~al.}(2014)\citenamefont{Luxmoore, Gan, Liu,
  Valmorra, Li, Faist, and Nash}}]{Luxmoore2014}
\bibinfo{author}{\bibfnamefont{I.~J.} \bibnamefont{Luxmoore}},
  \bibinfo{author}{\bibfnamefont{C.~H.} \bibnamefont{Gan}},
  \bibinfo{author}{\bibfnamefont{P.~Q.} \bibnamefont{Liu}},
  \bibinfo{author}{\bibfnamefont{F.}~\bibnamefont{Valmorra}},
  \bibinfo{author}{\bibfnamefont{P.}~\bibnamefont{Li}},
  \bibinfo{author}{\bibfnamefont{J.}~\bibnamefont{Faist}}, \bibnamefont{and}
  \bibinfo{author}{\bibfnamefont{G.~R.} \bibnamefont{Nash}},
  \bibinfo{journal}{ACS Photonics} \textbf{\bibinfo{volume}{1}},
  \bibinfo{pages}{1151} (\bibinfo{year}{2014}), ISSN \bibinfo{issn}{2330-4022},
  \urlprefix\url{http://pubs.acs.org/doi/abs/10.1021/ph500233s}.

\bibitem[{\citenamefont{Yan et~al.}(2013)\citenamefont{Yan, Low, Zhu, Wu,
  Freitag, Li, Guinea, Avouris, and Xia}}]{Yan2013}
\bibinfo{author}{\bibfnamefont{H.}~\bibnamefont{Yan}},
  \bibinfo{author}{\bibfnamefont{T.}~\bibnamefont{Low}},
  \bibinfo{author}{\bibfnamefont{W.}~\bibnamefont{Zhu}},
  \bibinfo{author}{\bibfnamefont{Y.}~\bibnamefont{Wu}},
  \bibinfo{author}{\bibfnamefont{M.}~\bibnamefont{Freitag}},
  \bibinfo{author}{\bibfnamefont{X.}~\bibnamefont{Li}},
  \bibinfo{author}{\bibfnamefont{F.}~\bibnamefont{Guinea}},
  \bibinfo{author}{\bibfnamefont{P.}~\bibnamefont{Avouris}}, \bibnamefont{and}
  \bibinfo{author}{\bibfnamefont{F.}~\bibnamefont{Xia}},
  \bibinfo{journal}{Nature Photonics} \textbf{\bibinfo{volume}{7}},
  \bibinfo{pages}{394} (\bibinfo{year}{2013}), ISSN \bibinfo{issn}{1749-4885},
  \urlprefix\url{http://www.nature.com/doifinder/10.1038/nphoton.2013.57}.

\bibitem[{\citenamefont{Tonouchi}(2007)}]{Tonouchi2007}
\bibinfo{author}{\bibfnamefont{M.}~\bibnamefont{Tonouchi}},
  \bibinfo{journal}{Nature Photonics} \textbf{\bibinfo{volume}{1}},
  \bibinfo{pages}{97} (\bibinfo{year}{2007}), ISSN \bibinfo{issn}{1749-4885}.

\bibitem[{\citenamefont{Woessner et~al.}(2014)\citenamefont{Woessner,
  Lundeberg, Gao, Principi, Alonso-Gonz\'{a}lez, Carrega, Watanabe, Taniguchi,
  Vignale, Polini et~al.}}]{Woessner2014a}
\bibinfo{author}{\bibfnamefont{A.}~\bibnamefont{Woessner}},
  \bibinfo{author}{\bibfnamefont{M.~B.} \bibnamefont{Lundeberg}},
  \bibinfo{author}{\bibfnamefont{Y.}~\bibnamefont{Gao}},
  \bibinfo{author}{\bibfnamefont{A.}~\bibnamefont{Principi}},
  \bibinfo{author}{\bibfnamefont{P.}~\bibnamefont{Alonso-Gonz\'{a}lez}},
  \bibinfo{author}{\bibfnamefont{M.}~\bibnamefont{Carrega}},
  \bibinfo{author}{\bibfnamefont{K.}~\bibnamefont{Watanabe}},
  \bibinfo{author}{\bibfnamefont{T.}~\bibnamefont{Taniguchi}},
  \bibinfo{author}{\bibfnamefont{G.}~\bibnamefont{Vignale}},
  \bibinfo{author}{\bibfnamefont{M.}~\bibnamefont{Polini}},
  \bibnamefont{et~al.}, \bibinfo{journal}{Nature Materials}
  \textbf{\bibinfo{volume}{14}}, \bibinfo{pages}{421} (\bibinfo{year}{2014}),
  ISSN \bibinfo{issn}{1476-1122},
  \urlprefix\url{http://www.nature.com/doifinder/10.1038/nmat4169}.

\bibitem[{\citenamefont{Lee et~al.}(2012)\citenamefont{Lee, Ahn, Shim, Lee, and
  Ryu}}]{Lee2012}
\bibinfo{author}{\bibfnamefont{J.~E.} \bibnamefont{Lee}},
  \bibinfo{author}{\bibfnamefont{G.}~\bibnamefont{Ahn}},
  \bibinfo{author}{\bibfnamefont{J.}~\bibnamefont{Shim}},
  \bibinfo{author}{\bibfnamefont{Y.~S.} \bibnamefont{Lee}}, \bibnamefont{and}
  \bibinfo{author}{\bibfnamefont{S.}~\bibnamefont{Ryu}},
  \bibinfo{journal}{Nature Communications} \textbf{\bibinfo{volume}{3}},
  \bibinfo{pages}{1024} (\bibinfo{year}{2012}), ISSN \bibinfo{issn}{2041-1723},
  \urlprefix\url{http://dx.doi.org/10.1038/ncomms2022}.

\end{thebibliography}

\begin{thebibliography}{1}
\expandafter\ifx\csname url\endcsname\relax
  \def\url#1{\texttt{#1}}\fi
\expandafter\ifx\csname urlprefix\endcsname\relax\def\urlprefix{URL }\fi
\providecommand{\bibinfo}[2]{#2}
\providecommand{\eprint}[2][]{\url{#2}}

\bibitem{Hale2011}
\bibinfo{author}{Hale, P.~J.}, \bibinfo{author}{Hornett, S.~M.},
  \bibinfo{author}{Moger, J.}, \bibinfo{author}{Horsell, D.~W.} \&
  \bibinfo{author}{Hendry, E.}
\newblock \bibinfo{title}{{Hot phonon decay in supported and suspended
  exfoliated graphene}}.
\newblock \emph{\bibinfo{journal}{Phys. Rev. B}} \textbf{\bibinfo{volume}{83}},
  \bibinfo{pages}{121404} (\bibinfo{year}{2011}).
\newblock \eprint{1012.3927}.

\bibitem{Tielrooij2013}
\bibinfo{author}{Tielrooij, K.~J.} \emph{et~al.}
\newblock \bibinfo{title}{{Photoexcitation cascade and multiple hot carrier
  generation in graphene}}.
\newblock \emph{\bibinfo{journal}{Nature Physics}}
  \textbf{\bibinfo{volume}{9}}, \bibinfo{pages}{248--252}
  (\bibinfo{year}{2013}).
\newblock \urlprefix\url{http://dx.doi.org/10.1038/nphys2564}.
\newblock \eprint{arXiv:1210.1205v2}.

\bibitem{Luxmoore2014b}
\bibinfo{author}{Luxmoore, I.~J.} \emph{et~al.}
\newblock \bibinfo{title}{{Strong Coupling in the Far-Infrared between Graphene
  Plasmons and the Surface Optical Phonons of Silicon Dioxide}}.
\newblock \emph{\bibinfo{journal}{ACS Photonics}} \textbf{\bibinfo{volume}{1}},
  \bibinfo{pages}{1151--1155} (\bibinfo{year}{2014}).
\newblock \urlprefix\url{http://pubs.acs.org/doi/abs/10.1021/ph500233s}.

\bibitem{kitamura2007}
\bibinfo{author}{Kitamura, R.}, \bibinfo{author}{Pilon, L.} \&
  \bibinfo{author}{Jonasz, M.}
\newblock \bibinfo{title}{{Optical constants of silica glass from extreme
  ultraviolet to far infrared at near room temperature.}}
\newblock \emph{\bibinfo{journal}{Applied optics}}
  \textbf{\bibinfo{volume}{46}}, \bibinfo{pages}{8118--8133}
  (\bibinfo{year}{2007}).

\bibitem{Wunsch2006}
\bibinfo{author}{Wunsch, B.}, \bibinfo{author}{Stauber, T.},
  \bibinfo{author}{Sols, F.} \& \bibinfo{author}{Guinea, F.}
\newblock \bibinfo{title}{{Dynamical polarization of graphene at finite
  doping}}.
\newblock \emph{\bibinfo{journal}{New Journal of Physics}}
  \textbf{\bibinfo{volume}{8}}, \bibinfo{pages}{318} (\bibinfo{year}{2006}).
\newblock \eprint{0610630}.

\end{thebibliography}

\clearpage


\setcounter{figure}{0}
\setcounter{equation}{0}
\setcounter{page}{1}

\renewcommand\thefigure{S\arabic{figure}}
\renewcommand\theequation{S\arabic{equation}}
\renewcommand{\bibnumfmt}[1]{[S#1]}
\renewcommand{\citenumfont}[1]{S#1}

\begin{center}
\textbf{\large{All-Optical Generation of Surface Plasmons in Graphene: Supplementary Information}}

\vspace{4mm}

T.~J.~Constant,$^{1,*}$ S.~M.~Hornett,$^1$ D.~E.~Chang$^{1,\dagger}$ and E.~Hendry$^1$

\textit{$^1$Electromagnetic Materials Group, Department of Physics, College of Engineering, Mathematics and Physical Sciences, University of Exeter, Exeter, Devon, UK. EX4 4QL.}

\textit{$^2$ICFO - Institut de Ci\`encies Fot\`oniques, Mediterranean
Technology Park, 08860 Castelldefels (Barcelona), Spain}
\end{center}

\section{Substrate Response}
\begin{figure}[b]
\begin{center}
\includegraphics[width=10cm]{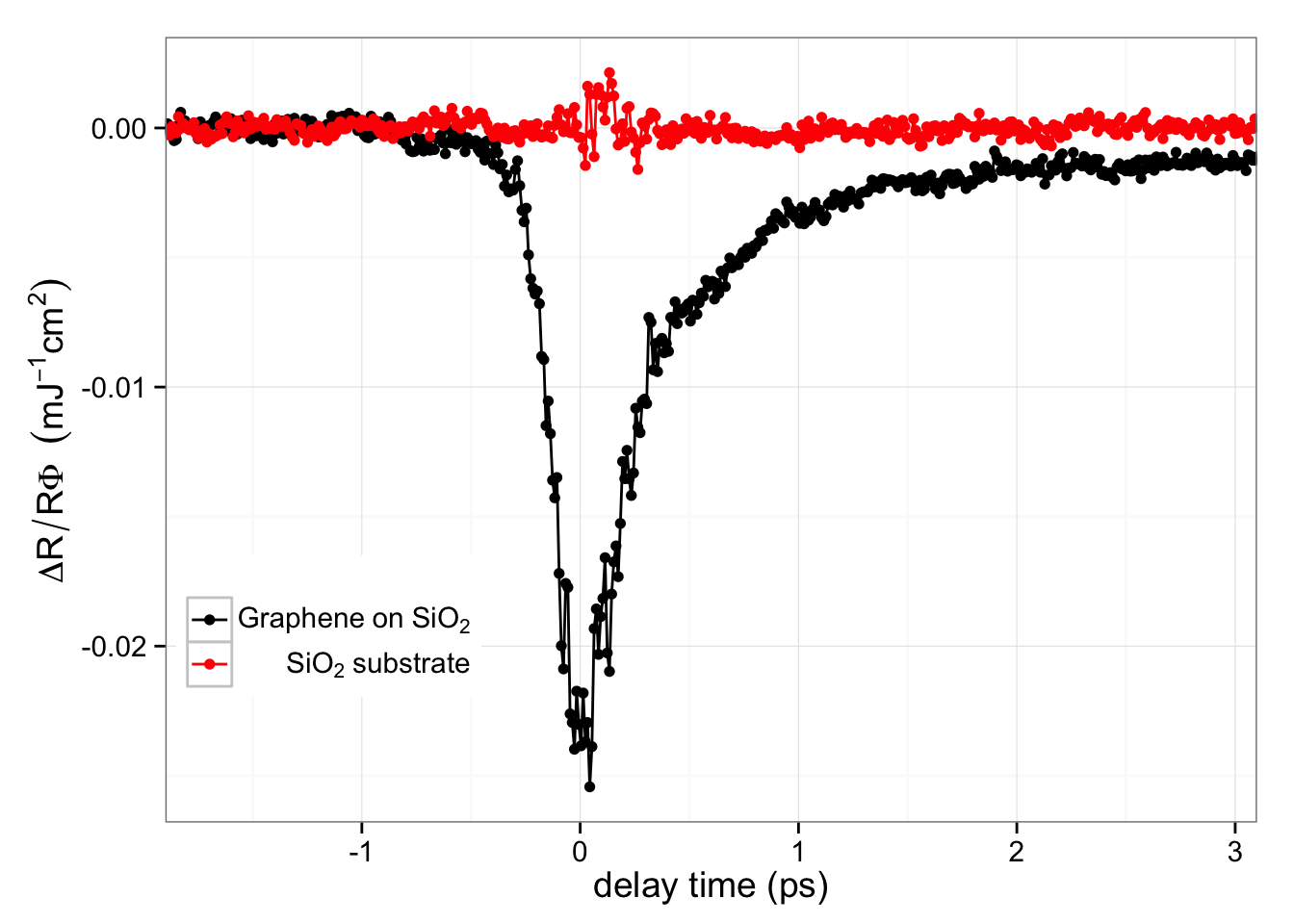}
\end{center}
\caption{Differential reflection normalized to fluence as a function of temporal overlap for: Graphene on quartz (black) and the bare quartz substrate (red) at $\theta_{pump}=50^\circ,~\theta_{probe}=70^\circ$. \label{fig:substrate_check}}
\end{figure}

Differential reflection was recorded as a function of delay time for both the graphene on quartz and for the bare quartz substrate to ascertain any contribution to the nonlinear response from the substrate. The experimental parameters for these data were set to the resonant condition of figure \ref{fig:dispersion_compilation}(b), with $\theta_{pump}=50^\circ,~\theta_{probe}=70^\circ$ and the difference frequency set to $10~\mathrm{THz}$. The measurements are compared in figure \ref{fig:substrate_check}, and negligible signal, compared to the resonant measurement from graphene, is observed. 

\section{Increasing Pump Fluence}
\begin{figure}[b]
\begin{center}
\includegraphics[width=15cm]{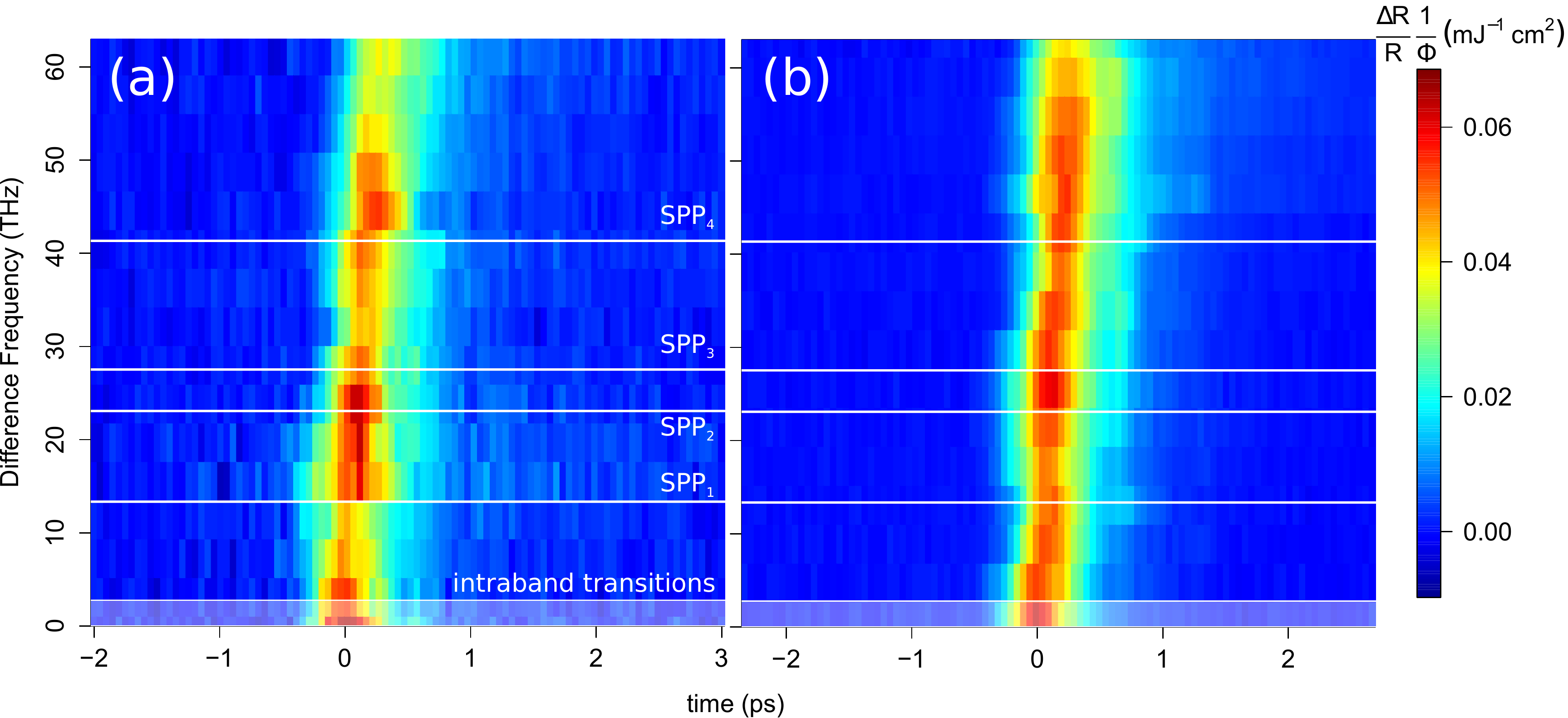}
\end{center}
\caption{Differential reflection normalized to fluence as a function of temporal overlap for: low (left) and high (right) pump fluence at $\theta_{pump} = 15^\circ$ and $\theta_{probe} = 125^\circ$. \label{fig:high_fluence}}
\end{figure}
For pulsed fluence $\sim 0.1~\mathrm{mJ/cm^2}$ we expect to generate an electron temperature of $\sim 1000~\mathrm{K}$\cite{Hale2011}. This means that we are probing a very non-equilibrium electron distribution. In order to investigate the effect of this electron heating, we increased the pump fluence used in the experiment to $\sim1.1~\mathrm{mJ/cm^2}$. These results are shown in fig.~\ref{fig:high_fluence}. We find that a higher fluence significantly suppresses the surface plasmon resonance features with respect to the background, off-resonance signal. We believe two factors contribute to this effect: firstly, an increased electron temperature will increase saturable absorption, the effect primarily responsible for the incoherent background signal. Secondly, due to the negative photoconductivity usually exhibited by graphene for pulsed femtosecond excitation\cite{Tielrooij2013}, one can expect increased losses and quenching of the surface plasmon, leading to broadening of the spectral features associated with their excitation.

Evidence that the increased electron temperature also raises the effective Fermi energy of the sample can be inferred by comparing fig.~\ref{fig:high_fluence}(a) and fig.~\ref{fig:high_fluence}(b), where the resonant regions of differential reflectivity shift to higher frequencies in the high-fluence case, as would be expected for the graphene surface plasmon dispersion for a higher doping level.

\section{Reducing Pump Fluence}
\begin{figure}[b]
\begin{center}
\includegraphics[width=15cm]{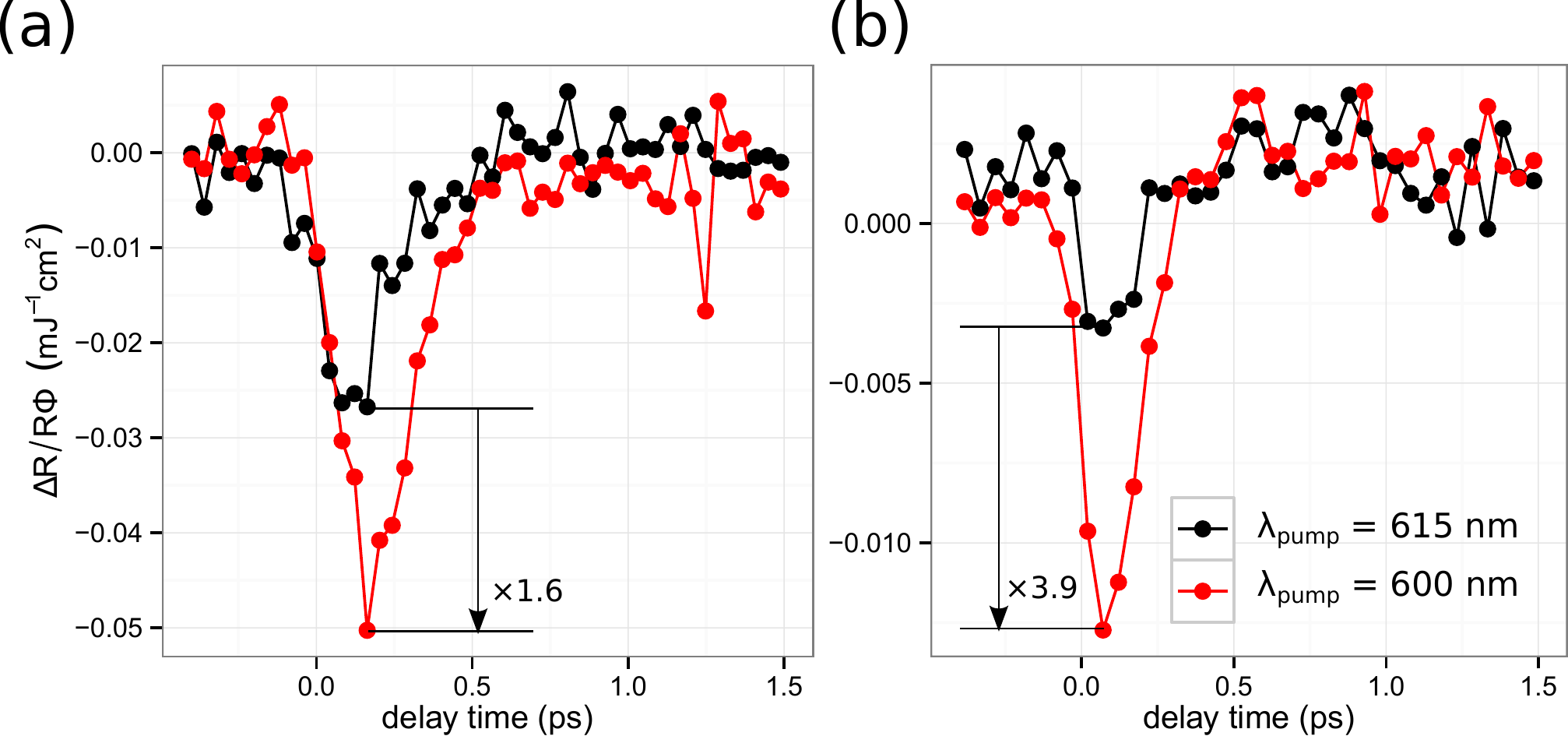}
\end{center}
\caption{Differential reflection normalized to pump fluence as a function of temporal overlap of the pulses for: (a) high-fluence pump and low-fluence probe, and (b) comparable fluence in both pump and probe. The black lines indicate the case for the difference frequency of 0~THz (no resonant plasmon coupling) and the red lines are for a difference frequency of 12~THz (resonant plasmon coupling). $\theta_{pump} = 55^\circ, \theta_{probe}= 70^\circ, \lambda_{probe} = 615~\mathrm{nm}$.\label{fig:low_fluence}}
\end{figure}
Since electron heating clearly has a negative impact in the experiment, one wants to minimize the pump intensity. However, reducing the light intensity also reduces our difference frequency coupling efficiency, since surface plasmon generation here is a nonlinear process. However, a possible route to better isolating coherent signals is to  reduce the pump beam intensity and increase the probe beam intensity, illuminating the sample with similar fluences for both beams. This reduces the Pauli blocking of the probe induced by the pump beam, decreasing the signal due to saturable absorption decreases, while maintaining the efficiency of the difference frequency mixing process.

Figures \ref{fig:low_fluence}(a) and \ref{fig:low_fluence}(b) show differential reflection normalized to pump fluence for two difference frequencies, 0~THz ($\lambda_{pump} = 615~\mathrm{nm}$) and 12~THz ($\lambda_{pump} = 600~\mathrm{nm}$), measured for the angles $\theta_{pump} = 50^\circ$ and $\theta_{probe} = 70^\circ$. In this geometry we expect a resonant enhancement for a difference frequency $\sim 12~\mathrm{THz}$ due to plasmon excitation and no enhancement for 0~THz, as previously measured in fig.~\ref{fig:dispersion_compilation}(b). In figure \ref{fig:low_fluence} (a) we show a typical measurement for a high-power pump beam ($0.26~\mathrm{mJ/cm^2}$) and a low power-probe beam ($0.0028~\mathrm{mJ/cm^2}$). In this case, when the difference frequency matches that of the surface plasmon, a resonant change to the reflectivity by a factor of 1.6 occurs. For equal pump and probe fluences ($\sim 0.07~\mathrm{mJ/cm^2}$), as shown in fig. \ref{fig:low_fluence}(b), we observe a significant suppression of the background, non-resonant signal. The enhancement then measured from a non-resonant condition to the resonant surface plasmon excitation is increased to a factor of 3.9. 

We also note that in the low-power pump case (fig. \ref{fig:low_fluence}(b)), the lineshape of the recorded temporal dynamics is far more symmetric than in the high-power case (fig. \ref{fig:low_fluence}(a)), which clearly exhibits the typical asymmetric lineshape indicative of incoherent carrier cooling dynamics.

\section{Higher Difference Frequencies}
\begin{figure}[ht]
\begin{center}
\includegraphics[width=12cm]{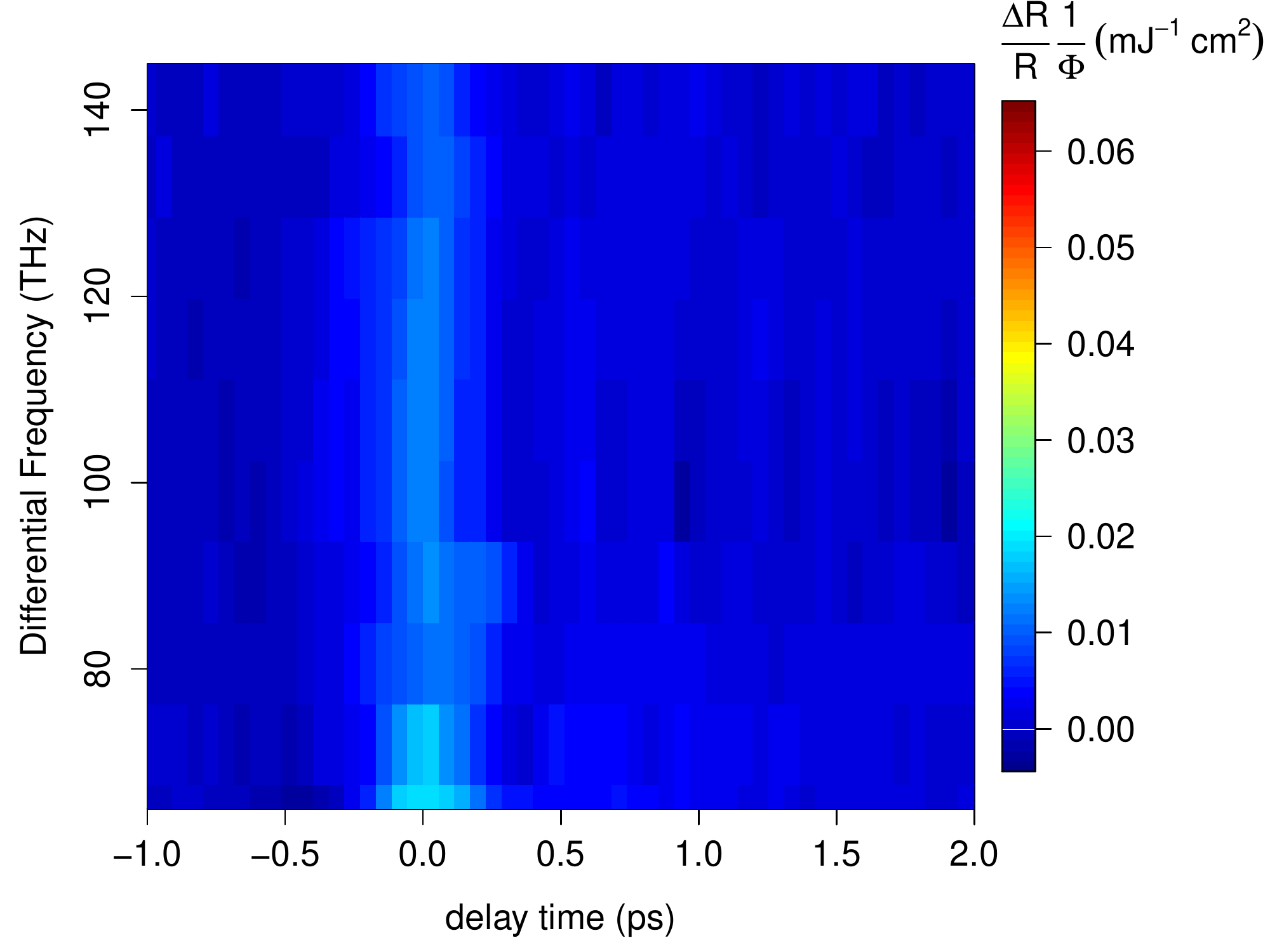}
\end{center}
\caption{Plot of differential reflection normalized to pump fluence as a function of temporal overlap for $\theta_{pump} = 15^\circ$ and $\theta_{probe} = 125^\circ$ at higher frequencies. The colorscale has been scaled to fig.~\ref{fig:dispersion_compilation}(c) for ease of comparison.\label{fig:high_freqs}}
\end{figure}
We have preformed several measurements for a wider range of difference frequencies. In figure \ref{fig:high_freqs}, we present measurements taken for $\theta_{pump} = 15^\circ$ and $\theta_{probe} = 125^\circ$, where the pump wavelength was varied from $540~\mathrm{nm}$ to $475~\mathrm{nm}$, with the probe wavelength fixed at $615~\mathrm{nm}$. This gives a difference frequency range from 70~THz to 140~THz. For these larger difference frequencies, we observe no resonance features above 70~THz, indicating there is no coherent coupling to higher frequency modes.

\section{Polarization Dependence}
\begin{figure}[t]
\begin{center}
\includegraphics[width=15cm]{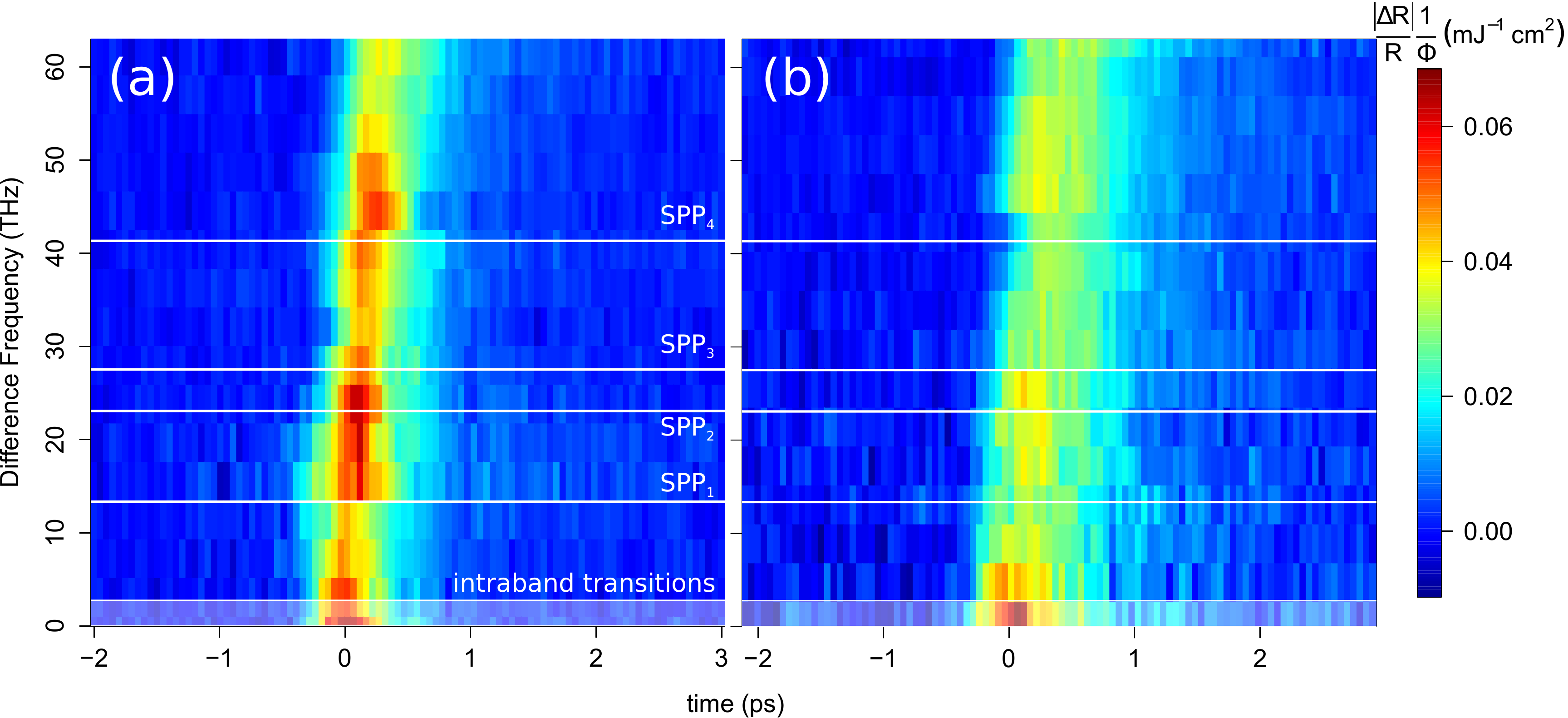}
\end{center}
\caption{Differential reflection normalized to pump fluence as a function of temporal overlap for transverse magnetic (a) and transverse electric (b) polarizations. The resonances for the predicted surface plasmon frequencies are clearly suppressed (labelled), while the intraband resonance is largely unaffected. \label{fig:polarisation}}
\end{figure}
The experiment shown in fig.~\ref{fig:dispersion_compilation}(c) was repeated with the pump and probe both polarized with the electric vector parallel to the graphene surface (transverse electric, TE polarized). Under these circumstances we expect surface plasmon excitation to be suppressed. The results are shown in figure \ref{fig:polarisation}. When illuminating with TE polarized light, there is a clear decrease in the nonlinear enhancement to the reflectivity for the 23-27~THz peak and the $\sim 45$~THz peak. The peak at $\sim0$~THz is of the same order in both polarization cases, suggesting that the nonlinearity arising due to the intraband transitions are less sensitive to the polarization of the light. This is attributed to how carriers respond to the EM field in the plane of the surface. Since, for the intraband response the conductivity functions are in-plane, they make no distinction as to the out of plane component (which distinguishes TE and TM polarization).

While the higher-frequency resonances are strongly suppressed, they are not absent completely: we attribute this to imperfect polarization of the beams and inhomogeneity in the graphene sample.

\section{Theoretical Model}
\begin{figure}[b]
\begin{center}
\includegraphics[width=12cm]{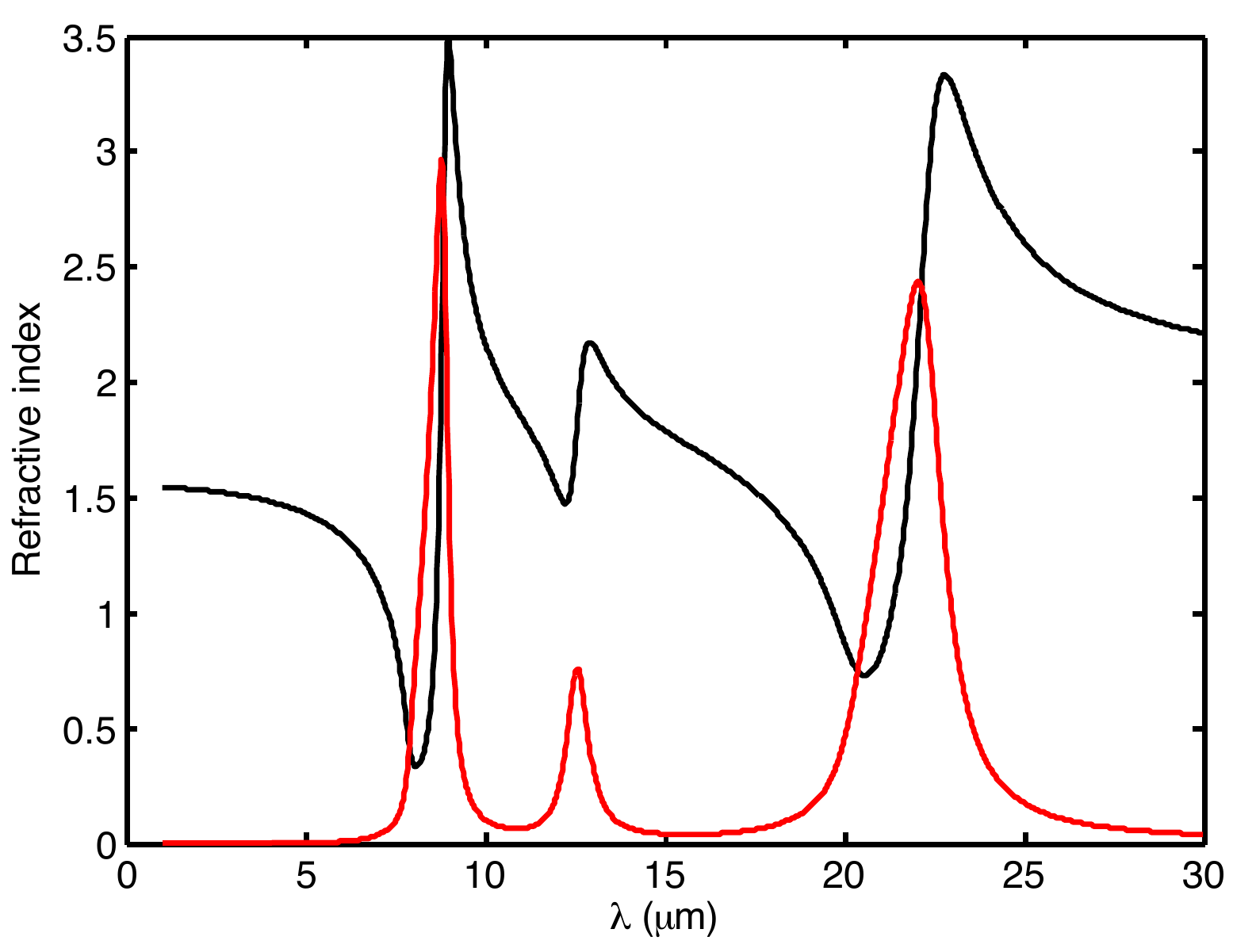}
\end{center}
\caption{Real~(black) and imaginary~(red) parts of the refractive index of silica versus free-space wavelength~($\lambda=2\pi c/\omega$), based upon the model of Eq.~(\ref{eq:n_SiO2}).\label{fig:SiO2model}}
\end{figure}

\begin{figure}[b]
\begin{center}
\includegraphics[width=12cm]{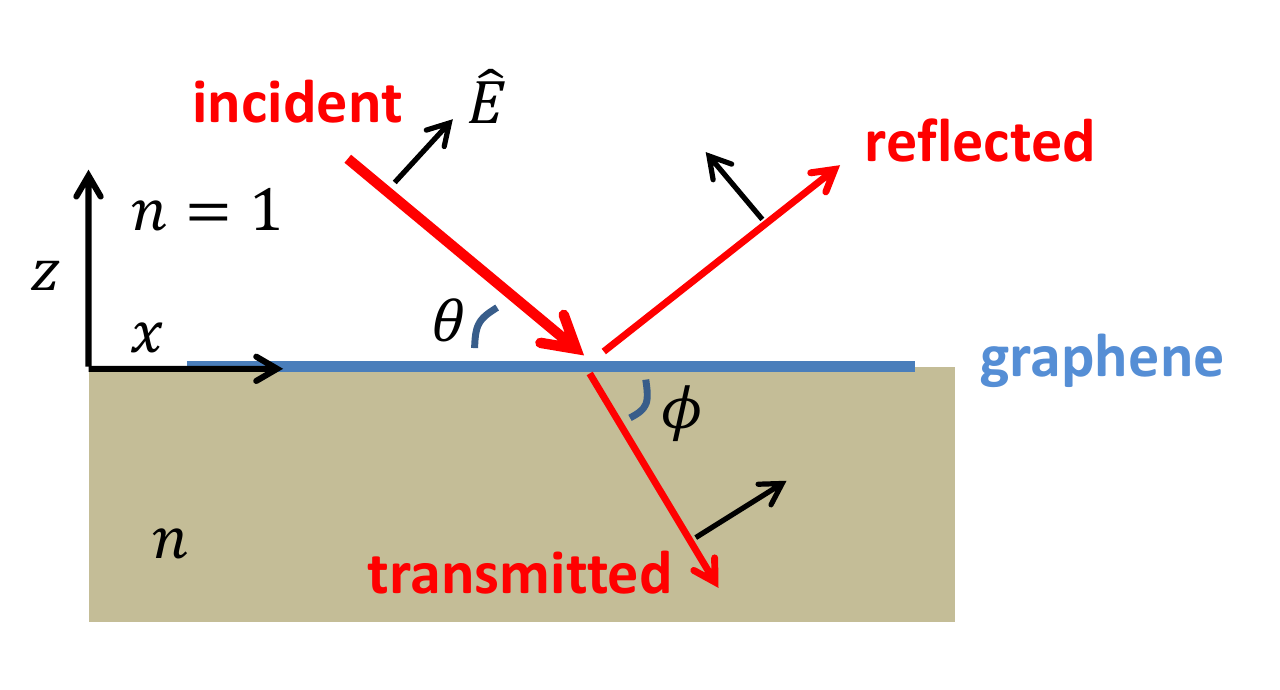}
\end{center}
\caption{Illustration of $p$-polarized electromagnetic fields propagating in the $x$-$z$ plane and interacting with graphene. The fields $i=1,2$ consist of incident, reflected, and transmitted components, with the directions of propagation and polarizations indicated by the red and black arrows, respectively. The angles of incidence and transmission are $\theta$ and $\phi$. The incident field is assumed to propagate in vacuum~(refractive index $n=1$), while the graphene sits on top of a substrate with index $n$~(possibly frequency dependent).\label{fig:convention}}
\end{figure}

We describe a theoretical model of two continuous-wave, free-space beams of frequencies $\omega_{1,2}$~(without loss of generality, assume that $\omega_1>\omega_2$) interacting with graphene via a difference frequency generation process. The convention in our calculations to define the field polarizations and beam angles is illustrated in Fig.~\ref{fig:convention}. The beams are taken to be incident from air (refractive index $n\approx 1$). Important to modeling this experiment is the inclusion of a frequency-dependent and complex refractive index of the substrate at low frequencies, in order to capture the lattice vibrations in silica and the resulting surface optical phonons. To do this, we take a simple dielectric response model based upon three transverse optical~(TO) phonon modes~\cite{Luxmoore2014b},
\be n^2(\omega)=\epsilon_{\infty}+\sum_{j=1}^{3}\frac{f_j\omega_{TO,j}^2}{\omega_{TO,j}^2-\omega^2-i\omega\gamma_{TO,j}}.\label{eq:n_SiO2} \ee
From Ref.~\cite{Luxmoore2014b}, the high-frequency dielectric constant is taken to be $\epsilon_{\infty}=2.4$, while the TO phonon frequencies and oscillator weights are $\omega_{TO}=2\pi\times(13.44, 23.75, 33.84)$~THz and $f=(0.7514,0.1503,0.6011)$, respectively. The damping rates are taken to be $\gamma_{TO}=2\pi\times(0.80, 1.27, 1.27)$~THz. The resulting real and imaginary parts of the refractive index, plotted in Fig.~\ref{fig:SiO2model}, approximately correspond to experimentally measured values~\cite{kitamura2007}. In practice, this refractive index function is only relevant for the substrate response at the low difference frequency of $\omega_3=\omega_1-\omega_2$, while for the high frequencies $\omega_{1,2}$ the response is nearly frequency-independent, $n\approx \sqrt{\epsilon_{\infty}}$.

In general, one can obtain equations relating the reflection and transmission coefficients to each other by enforcing electromagnetic boundary conditions (continuity of the normal electric displacement and tangential electric field) at the graphene interface. The solution for the transmission coefficient of field $i$~($i=1,2$) is readily found to be
\be t_i=\frac{2\sin\theta_i}{n_i\sin\theta_i+\sin\phi_i}-\frac{(\rho_{is}/\epsilon_0)\sec\phi_i\sin\theta_i}{E_{Ii}(n_i^2\sin\theta_i+n_i\sin\phi_i)},\label{eq:t}
\ee
where $n_i=n(\omega_i)$ denotes the substrate refractive index at the field frequency, while the reflection coefficient is related by $r_i=1-t_i \sin\phi_i \csc\theta_i$. Here $E_{Ii}$ are the incident field amplitudes, and $\theta_i$ and $\phi_i$ are the angles of the fields on the vacuum and substrate sides, respectively. $\rho_{is}=\rho(\omega_i,k_{ix})$ is the graphene surface charge density at frequency $\omega_i$ and in-plane wavevector $k_{ix}=(\omega_i/c)\cos\theta_i$. Note that the first term on the right-hand side of Eq.~(\ref{eq:t}) reproduces the standard Fresnel coefficient in the absence of a graphene layer~($\rho_{is}=0$). The angle of the transmitted field is related to the incident by Snell's Law, $\cos\theta_i=n_i \cos\phi_i$.

The surface charge density can be related to the current density $J$ in the graphene layer via the continuity equation, which in the Fourier domain reads
\be \rho_s(\omega,k_x)=(k_x/\omega)J_x(\omega,k_x).\label{eq:continuity} \ee
At the same time, the current density can be related to the electric fields via conductivity functions. We are particularly interested in the case of difference frequency generation, where the field produced at the difference frequency and wavevector $\omega_3=\omega_1-\omega_2$ and $k_{3x}=k_{1x}-k_{2x}$ is resonantly enhanced by aligning them with the plasmon dispersion of graphene $\omega_p(k_x)$. This motivates a truncated model in which we include only the linear and second-order conductivities, and frequencies $\omega_{1,2}$ and $\omega_3$~(thus left out is sum frequency generation and the generation of even higher harmonics). Then, the current density for field $1$ is given by
\be J_{x}(\omega_1,k_{1x})=\sigma^{(1)}(\omega_1,k_{1x})E_x(\omega_1,k_{1x})+\sigma^{(2)}(\omega_1,k_{1x};\omega_2,k_{2x},\omega_3,k_{3x})E_x(\omega_2,k_{2x})E_x(\omega_3,k_{3x}).\label{eq:J1x} \ee
Here, $E_{x}(\omega_1,k_{1x})=t_1 E_{1I} \sin\phi_{1}$ is the total parallel field for $i=1$ at the graphene layer, which we have written in terms of the incident field and transmission coefficient. $\sigma^{(1)}$ is the linear conductivity function, while $\sigma^{(2)}(\omega_1,k_{1x};\omega_2,k_{2x},\omega_3,k_{3x})$ is the second-order nonlinear conductivity functions relating the current density generated at $\omega_1,k_{1x}$ given fields at $\omega_2,k_{2x}$ and $\omega_3,k_{3x}$. Similar expressions as Eq.~(\ref{eq:J1x}) can be written down for the current density at $\omega_i,k_{ix}$~($i=2,3$), and we use an analogous set of conventions to indicate the fields and conductivities at other frequencies and wavevectors. In what follows, we will also adopt the more compact notation $\sigma^{(2)}(\omega_1)=\sigma^{(2)}(\omega_1,k_{1x};\omega_2,k_{2x},\omega_3,k_{3x})$, where the dependence on wavevectors and input frequencies is understood.

The substitution of Eqs.~(\ref{eq:continuity}) and~(\ref{eq:J1x}) into Eq.~(\ref{eq:t})~(along with analogous equations for the other fields $i=2,3$) yields a set of nonlinear equations relating the transmission coefficients and incident fields,
\bea t_1 & = & t_{1}^{(L)}\left[1-\frac{|t_2 E_{2I}|^2}{(2c\epsilon_0)^2}t_{1}^{(L)}t_{3}^{(L)}\sigma^{(2)}(\omega_1)\sigma^{(2)}(\omega_3)\sin\phi_1\sin^2\phi_2\sin\phi_3\right]^{-1},\label{eq:t1} \\ t_2 & = & t_{2}^{(L)}\left[1-\frac{|t_1 E_{1I}|^2}{(2c\epsilon_0)^2}t_{2}^{(L)}t_{3}^{(L)*}\sigma^{(2)}(\omega_2)\sigma^{(2)*}(\omega_3)\sin^2\phi_1\sin\phi_2\sin^{\ast}\phi_3\right]^{-1}.\label{eq:t2} \eea
Here $t^{(L)}=\frac{2\sin\theta}{n\sin\theta+\sin\phi+(\sigma^{(1)}/c\epsilon_0)\sin\theta\sin\phi}$ is the linear transmission coefficient, and the angle of the generated field is defined via $k_{3x}\equiv \frac{n_3\omega_3}{c}\cos\phi_3$. The complex in-plane field amplitude generated at the difference frequency and at the position of the graphene layer $z=0$ is given by
\be
E_{3x}=-\frac{t_1t^{\ast}_2t_3^{(L)}}{2c\epsilon_0}E_{1I}E_{2I}^{\ast}\sigma^{(2)}(\omega_3)\sin\phi_1\sin\phi_2\sin\phi_3.\label{eq:E3} \ee

While Eqs.~(\ref{eq:t1}) and~(\ref{eq:t2}) may appear somewhat complicated, here we note their main features. First, we note that the input field amplitudes, the beam angles, and the linear optical properties of the system are generally known. Thus, on one hand, given a theoretical model of the nonlinear conductivity $\sigma^{(2)}$, these equations can be solved to obtain the predicted changes in transmission and reflection of the input beams, due to the generation of plasmons at $\omega_3,k_{3x}$. The plasmon field amplitude itself can be found from Eq.~(\ref{eq:E3}). On the other hand, even absent a theoretical model, if changes in transmission or reflection of the incident fields are experimentally measured, one can attempt to invert these equations in order to obtain an experimentally inferred value of $\sigma^{(2)}$.

The description above generally holds regardless of the values of $\omega_3,k_{3x}$. It is particularly interesting, however, to focus on the case where they align with the plasmon dispersion relation. In the small wavevector limit and for frequencies smaller than twice the Fermi frequency, $\omega\lesssim 2\omega_F$, the linear conductivity is well-approximated by the Drude model~\cite{Wunsch2006},
\be \sigma(\omega)\approx \frac{ie^2}{\pi\hbar}\frac{\omega_F}{\omega+i\gamma}. \ee
Here we have included a phenomenological damping term $\gamma$. The plasmon dispersion relation can be found by solving for the pole of the linear transmission coefficient, $t^{(L)}$. For simplicity, we will momentarily consider the case of a substrate with frequency-independent refractive index, so that the role of plasmon damping can be more clearly identified. In the absence of losses, the dispersion relation is found to be
\be k_{x,p}=\frac{(1+n^2)\omega_p^2}{4\alpha c\omega_F}, \ee
where $\alpha\approx 1/137$ is the fine-structure constant. In the presence of losses, choosing $\omega_3,k_{3x}$ to lie on the plasmon dispersion relation yields a linear transmission amplitude of $|t_3^{(L)}|\approx\frac{2nQ}{1+n^2}$, where $Q=\omega_3/\gamma$ is the plasmon quality factor. Under these conditions, one thus sees from Eq.~(\ref{eq:E3}) that the field intensity experiences a resonant enhancement of $|E_3|^2\propto Q^2$. A similar resonant effect appears in the transmission and reflection coefficients of the incident fields.

While the Eqs.~(\ref{eq:t1}) and~(\ref{eq:t2}) can in principle be inverted to infer $\sigma^{(2)}$ given experimental data for reflection or transmission coefficients, in the present experimental setup this procedure can only be done semi-quantitatively due to a number of unknowns. First, the signal lies significantly above the noise floor only near the plasmon dispersion relation, and only a limited number of beam angles are investigated. This makes it difficult to infer a specific wavevector and frequency dependence of the nonlinear conductivity~(fundamentally, there must be a dependence on wavevector, as otherwise $\sigma^{(2)}=0$ for a centrosymmetric material). Furthermore, the experiment employs pulses whose bandwidths are significantly larger than the plasmon linewidth. Given that, here we aim to reach a conservative estimate for the strength of $\sigma^{(2)}$, while we anticipate that future improved experiments~(such as with longer pulses and nano-structures) and theoretical models will enable more detailed comparisons.

The full conductivity function at zero temperature is given by~\cite{Wunsch2006}
\be \sigma(\omega)= \frac{ie^2}{\pi\hbar}\frac{\omega_F}{\omega+i\gamma}+\frac{e^2}{4\hbar}\left[\Theta(\omega-2\omega_F)+\frac{i}{\pi}\log \left|\frac{\omega-2\omega_F}{\omega+2\omega_F}\right|\right], \ee
where $\Theta(x)$ is the Heaviside step function. The Fermi energy $\hbar\omega_F\approx 0.5$~eV of graphene is significantly lower than the pump and probe photon energies of $\sim 3$~eV. At these frequencies, the linear conductivity of graphene is nearly frequency independent and real, $\sigma^{(1)}(\omega)/(c\epsilon_0)\approx \pi\alpha$, which we use to obtain the pump and probe linear reflection coefficients. Furthermore, we take the simplest possible function for the nonlinear conductivity, $\sigma^{(2)}(\omega)=i|\sigma^{(2)}(\omega_2)|(\omega/\omega_2)$, where $|\sigma^{(2)}(\omega_2)|$ is a single fitting parameter~(the probe frequency $\omega_2$ is fixed in the experiment). With this choice of function, graphene would be equivalent to a nonlinear material with a frequency-independent bulk nonlinear susceptibility of $\chi^{(2)}=-i\sigma^{(2)}(\omega)/(\omega\epsilon_0 t)=|\sigma^{(2)}(\omega_2)|/(\omega_2 \epsilon_0 t)$, where $t$ is the effective thickness of graphene. Inserting this nonlinear conductivity into Eqs.~(\ref{eq:t1}),~(\ref{eq:t2}), and~(\ref{eq:E3}), we find that a value of $|{\sigma}^{(2)}(\omega_2)|\approx 2.4 \times 10^{-12}$~A$\cdot$m/V${}^2$ produces a good qualitative fit to the experimental data.
\begin{figure}[b]
\begin{center}
\includegraphics[width=15cm]{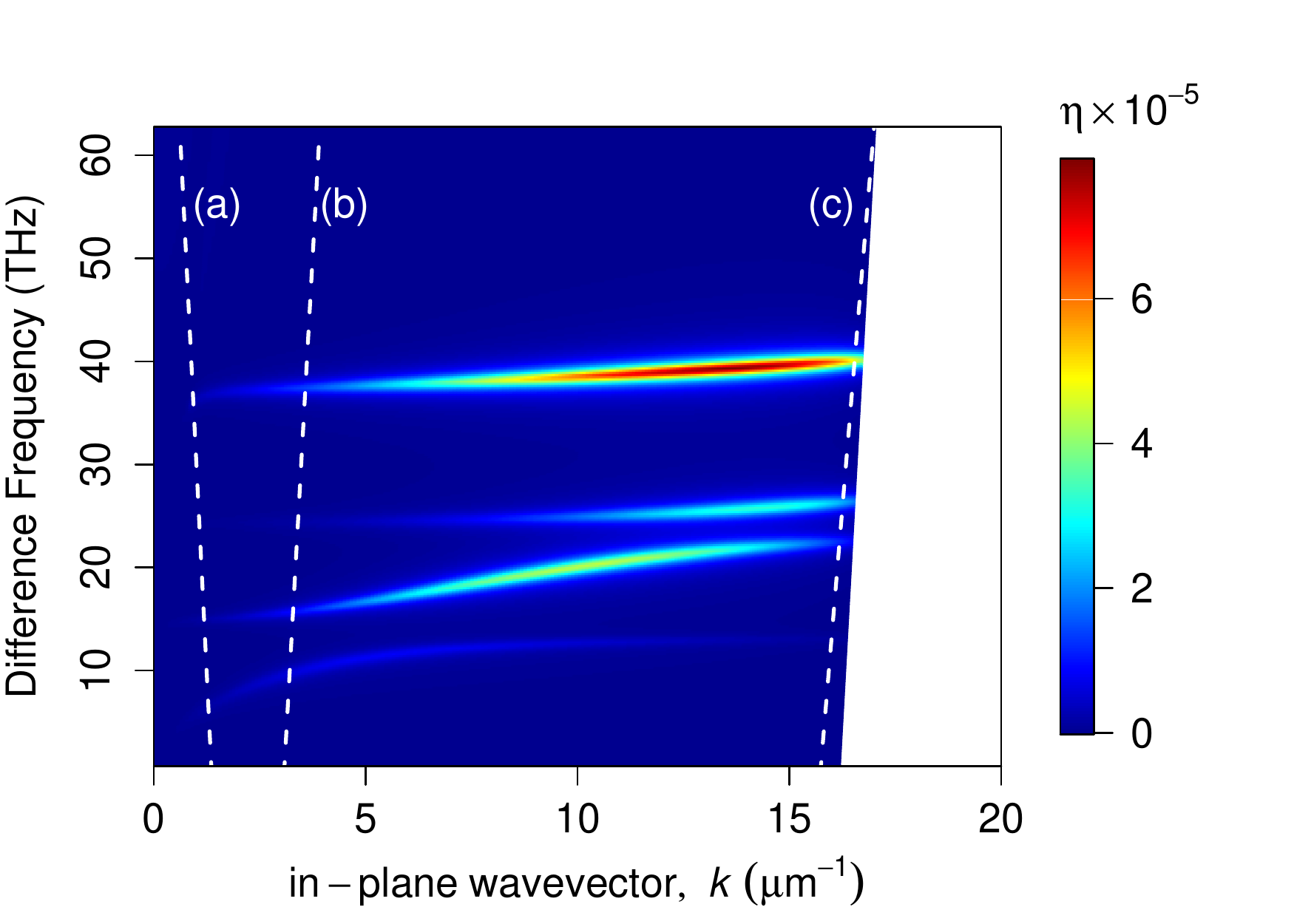}
\end{center}
\caption{The numerical solution for the pump conversion efficency, $\eta$, as a function of wavevector and difference frequency.\label{fig:theory_eta}}
\end{figure}

We now discuss how to obtain the conversion efficiency of pump photons to plasmons. The number of photons dissipated per unit area and time by the field at the difference frequency consists of two terms, $\Gamma_d=\Gamma_{d,g}+\Gamma_{d,s}$. The first term consists of damping from the graphene layer due to the real part of its conductivity, and is given by $\Gamma_{d,g}=\left(\textrm{Re}\;\sigma^{(1)}(\omega_3)\right)|E_{3x}|^2/(2\hbar\omega_3)$. The second term is due to damping from the substrate, due to the fact that at low frequencies its refractive index is complex. This contribution is given by $\Gamma_{d,s}=\frac{\epsilon_0}{4\hbar |\textrm{Im}\;(k_{3x}\tan\phi_3)|}(\textrm{Im}\;n^2(\omega_3))|E_{3x}|^2(1+|\cot\phi_3|^2)$. On the other hand, the incident photon flux in the pump field is $\Gamma_{\footnotesize\textrm{in}}=I_{1}\sin\theta_{1}/(\hbar\omega_1)$, where $I_1$ is the pump intensity. Generally, the amplitude of the generated plasmon field rate will depend on both the pump and probe intensities. However, at the level of individual photons, the process is that an incoming pump photon gets converted to a plasmon (assisted by stimulated emission of a photon into the probe). Thus, we define the conversion efficiency relative to the pump alone. In steady state, the rates of photons dissipated and generated at the difference frequency are equal, and thus the overall conversion efficiency of pump photons to plasmons is $\eta=\Gamma_{d}/\Gamma_{\footnotesize\textrm{in}}$. This efficiency is shown in fig.~\ref{fig:theory_eta} for the range of frequencies and wavevectors used in our experiments. Using the estimated value of $|{\sigma}^{(2)}(\omega_2)|$, we find that the conversion efficiency for the experimental arrangement shown in fig.~\ref{fig:dispersion_compilation}(b) at the point of maximum signal is approximately $\eta\approx 6 \times 10^{-6}$.

\end{document}